\documentclass[english,pra,groupedaddress,showpacs,twocolumn]{revtex4}
\usepackage{times}
\usepackage[T1]{fontenc}
\usepackage[latin1]{inputenc}
\usepackage{verbatim}
\usepackage{amsmath}
\usepackage{graphicx}
\usepackage{amssymb}

\makeatletter
\usepackage{babel}
\makeatother
\begin{document}

\title{Bistability and mode interaction in microlasers}

\author{Sergei V. Zhukovsky}

\altaffiliation{Present address: Institute of High-Frequency and Communication Technology, Faculty of Electrical, Information, and Media Engineering, University of Wuppertal, Rainer-Gruenter-Str. 21, D-42119 Wuppertal, Germany}

\affiliation{Physikalisches Institut and Bethe Center for Theoretical Physics,
Universit\"at Bonn, Nussallee 12, D-53115 Bonn, Germany}

\email{sergei@uni-wuppertal.de}

\author{Dmitry N. Chigrin}

\altaffiliation{Present address: Institute of High-Frequency and Communication Technology, Faculty of Electrical, Information, and Media Engineering, University of Wuppertal, Rainer-Gruenter-Str. 21, D-42119 Wuppertal, Germany}

\affiliation{Physikalisches Institut and Bethe Center for Theoretical Physics,
Universit\"at Bonn, Nussallee 12, D-53115 Bonn, Germany}

\author{Johann Kroha}

\affiliation{Physikalisches Institut and Bethe Center for Theoretical Physics,
Universit\"at Bonn, Nussallee 12, D-53115 Bonn, Germany}

\begin{abstract}
We investigate the possibility of bistable lasing in microcavity lasers
as opposed to bulk lasers. To that end, the dynamic behavior of a
microlaser featuring two coupled, interacting modes is analytically
investigated within the framework of a semiclassical laser model,
suitable for a wide range of cavity shapes and mode geometries. Closed-form
coupled mode equations are obtained for all classes of laser dynamics.
We show that bistable operation is possible in all of these classes.
In the simplest case (class-A lasers) bistability is shown to result
from an interplay between coherent (population-pulsation) and incoherent
(hole-burning) processes of mode interaction. We find that microcavities
offer better conditions to achieve bistable lasing than bulk cavities,
especially if the modes are not degenerate in frequency. This results
from better matching of the spatial intensity distribution of microcavity
modes. In more complicated laser models (class-B and class-C) bistability
is shown to persist for modes even further apart in frequency than
in the class-A case. The predictions of the coupled mode theory have
been confirmed using numerical finite-difference time-domain calculations.
\end{abstract}


\maketitle

\section{Introduction\label{sec:INTRO}}

In recent years, microlasers have been an object of growing interest
in the photonics community because of a remarkable promise in both
basic and applied research. Modern technology has facilitated fabrication
of high-$Q$ micro- and nanosized cavities (microresonators) in a
vast variety of designs (microdisks, -rings, -gears, -toroids, nanowires,
nanoposts, and so on \cite{VahalaReview}). Lasers can be based on
many of these set-ups as well as on different materials, e.g., semiconductors,
impurity ions, or dye molecules. In addition, periodic nanostructures
(photonic crystals, PhCs) can provide both cavity-based and distributed
feedback resonators suitable for laser design \cite{PBGdefect,PBGlaser}.
The cavity size, which becomes so small as to be comparable to the
operating wavelength, is what makes a microlaser physically distinct
from conventional ({}``bulk'') cavities whose size is far larger.
The small size limits the number of cavity modes that could take part
in lasing, and at the same time greatly increases the influence of
the cavity shape on the character of the modes. As a result, the mode
structure becomes more complicated and heavily dependent on the specific
cavity design. One is no longer able to describe the modes universally
in an analytical manner. The variety of laser dynamics becomes much
richer, which complicates the studies of microlasers to a considerable
extent but at the same time can harbor interesting new effects. For
example, one could look for new possibilities of bistable or multistable
lasing \cite{BabaDisks}, which would prove useful in many applications
such as multiple-wavelength light sources, optical flip-flop devices
or optical memory cells \cite{NatRings}.

In the simplest case when two modes coexist in the same laser cavity
(competing for the same saturable gain medium), three lasing regimes
are usually considered \cite{Siegman}. First, when one of the modes
has an advantage (e.g., larger $Q$-factor or better coupling to the
gain), it simply dominates, becoming the only lasing mode (\emph{single
mode lasing}). Second, when the modes are well balanced (i.e., similar
$Q$-factors and equally well coupled to the gain), they can both
lase simultaneously. Such a coexistence can become possible because
the modes with different frequency and/or spatial field pattern preferably
interact with different gain centers. As a consequence, the spectral
and spatial hole burning causes each mode to get saturated independently
and allow the mode that happens to be weaker to catch up with the
stronger one. Each mode saturates itself more readily as it does the
other mode; in this sense, the coexisting modes are said to be \emph{weakly
coupled} (\emph{simultaneous multimode lasing}). Third, if the reverse
is true, i.e., if each mode saturates the other mode before coming
to its own saturation (the modes are \emph{strongly coupled}), the
weaker mode is quenched by the stronger one before it has any chance
to catch up. Whichever mode has an initial advantage wins the competition
and becomes the only lasing mode (\emph{bistable multimode lasing}).
The system can lase in either mode and is in this sense bistable.

Trying to understand the physical origin of strong mode coupling brings
about certain problems. It was pointed out from the beginning \cite{LambLaser}
that harmonic modes (such as longitudinal modes in bulk cavities)
must always be coupled weakly because the antinodes of the field (the
regions where the light-matter interaction is maximized) are spatially
mismatched for different modes. Spatial hole burning would work similarly
for any two modes with mismatched intensity distribution (such as
transverse modes in bulk cavities). One of the ways to circumvent
this limitation is to use degenerate modes with identical spatial
intensity profiles, e.g., polarization degenerate modes or counterpropagating
modes in ring lasers. This can make the lasing bistable due to additional
mode coupling through population pulsations \cite{diRing,Siegman}.
Alternatively, one can place a saturable absorber in addition to the
saturable gain medium into the cavity \cite{mcSatAbsorb,mcSatAbsorb2,LambLaser}.
Such an absorber can be naturally realized when only a part of the
active medium is pumped. Both principles can be adapted for use in
microlasers and are embodied in the form of polarization-bistable
and absorptive bistable laser diodes \cite{diPBLD}. It has also been
shown that two coupled lasers can achieve bistability if the output
from each laser is directed to the other one and the feedback is reduced
to prevent formation of a compound cavity \cite{mcCoupledAgrawal,mcCoupledOudar}.
Later studies \cite{mcWieczorekOC,mcWieczorekPRA} give a detailed
account on the stability and mode locking regimes of bulk coupled
lasers based on nonlinear bifurcation analysis of the corresponding
rate equations. It is fundamentally problematic to achieve similar
behavior in microlasers where the modes share the same cavity. Recent
achievements in the design of bistable multimode-interference laser
diodes \cite{diMMI-BLD}, though capable of bistable lasing within
a cavity of sub-millimeter size, still require saturable absorbers
for the device to function properly. 

In the meantime, recent results show that there are yet unexplored
possibilities for bistable operation of microlasers. We have shown
\cite{zhukPRL} that a cavity based on coupled defects in a PhC exhibits
bistability without the need for saturable absorption or similar additional
mechanisms. %
\begin{comment}
The device can be brought into the bistable regime by properly choosing
 the mode frequencies with respect to the gain line profile. 
\end{comment}
{}The same idea was seen to work in lasers based on multimode nanopillar
waveguides \cite{zhukPSS}. Similar results have been reported based
on coupled microdisk \cite{BabaDisks} and coupled microring \cite{NatRings}
resonators, the latter proposed for an ultrafast, ultralow-power optical
memory cell design. Also, Ref.~\cite{diRingFlipFlop} reports that
coupled multiple-feedback ring lasers can be brought to bistability
by carefully selecting the feedback times, which may be more feasible
in microlasers than the conventional gain-quenching scheme as in \cite{mcCoupledOudar}.
Finally, a time-independent multimode laser theory recently developed
by H.~T\"{u}reci and co-workers \cite{Tureci06,Tureci07} reports
that mode interaction can be very important in highly multimode nanostructure-based
systems such as random lasers \cite{TureciScience}. In view of this,
there is a pronounced need to address the question of bistability
in microresonators with their specific features such as complex cavity
shapes and mathematically complex cavity modes taken into account
consistently. Spatial hole burning should also be accounted for rigorously
without reverting to averaging approximations, which are usually applied
for coupled or semiconductor lasers \cite{BabaDisks,mcCoupledAgrawal,mcWieczorekPRA}. 

In this paper, we consider the dynamics of two interacting modes in
a microresonator-based laser. The semiclassical rate-equation model
based on the Maxwell-Bloch equations is used to model a laser-active
medium. Coupled mode equations are derived and analyzed for different
classes of laser dynamics. Compared to existing accounts on mode dynamics
and coupled lasers \cite{mcHodges,mcWieczorekOC,mcWieczorekPRA,mcJohnBusch},
no specific form is assumed for either the cavity or the mode geometry.
The spatial distribution of population inversion is taken into account
fully in terms of projections onto the modes' subspace (see \cite{zhukFDTD})
for all classes of laser dynamics. The theory developed here can be
seen as complementary to the account in Refs.~\cite{Tureci06,Tureci07}
by being able to provide a description of time-dependent laser dynamics.
Though they are rather different, both these approaches go beyond
the third-order nonlinearity in the description of light-matter interaction.

In the simplest case of class-A laser dynamics, the equations suitable
for analytical studies have been derived. As already shown earlier
for some particular cases (see, e.g., \cite{mcWieczorekPRA}), we
confirm that coherent mode interaction (population pulsations) can
result in bistable laser operation. We show that bistable lasing becomes
increasingly more difficult to achieve as the intermode frequency
spacing~$\Delta\omega$ increases from zero. However, for microcavity
modes with well-matched intensity-gain overlap the bistability window
has been shown to be much greater (by up to several orders of magnitude
with respect to~$\Delta\omega$) than for harmonic bulk-cavity modes.
A non-symmetric system, where one of the modes is given an advantage
through cavity design, is also investigated. We show that a parameter
mismatch favoring one of the modes can be compensated for by an opposing
mismatch in another parameter that would favor the other mode. In
the more complicated class-B or class-C cases, numerical studies of
the obtained coupled mode equations have been carried out. The effects
of increasing the pumping rate and/or~$\Delta\omega$ beyond the
applicability limits of the class-A approximations are studied. Bistable
lasing is seen to persist unless $\Delta\omega$~becomes comparable
to the width of the gain line. Even then, bistability can be further
restored by increasing the pumping rate highly above threshold. The
results obtained for the class-B/C microlaser systems in the framework
of the coupled mode theory have been compared with full numerical
finite difference time domain (FDTD) calculations. At least for the
system considered (coupled defects in a 2D photonic crystal as in
Ref.~\cite{zhukPRL}), we demonstrate that the predictions of the
theory are in a good agreement with the results of numerical simulations.

The paper is structured as follows. In Section~\ref{sec:EQUATIONS},
we derive the semiclassical coupled two-mode laser equations suitable
for a wide range of microcavity modes. Only a few general assumptions
about the cavity shape are made and no particular form for the mode
geometry is specified. The derivation starts from the Maxwell-Bloch
equations and is carried out from the more general (class-C) through
the intermediate (class-B) to the most restrictive (class-A) laser
dynamics. Specific issues pertaining to introducing the dynamics classes
in multimode lasers are addressed along the way. The analysis of the
equations obtained is then carried out in the reverse order. In Section~\ref{sec:CLASS-A},
we analyze the class-A case, which, with some assumptions, turns out
to be closely related to the standard two-mode competition model \cite{Siegman}.
The parameter window of bistable operation is investigated in terms
of the spatial and spectral mode properties. In Section~\ref{sec:CLASS-BC},
class-B and class-C equations are numerically investigated, and the
main differences with the class-A model as regards bistable lasing
operation are discussed. Finally, Section~\ref{sec:SUMMARY} summarizes
the paper.

\section{Coupled two-mode laser equations\label{sec:EQUATIONS}}

\subsection{Semiclassical laser equations and multimode expansion\label{sub:EQS-GEN}}

\begin{figure}
\includegraphics[width=0.3\textwidth]{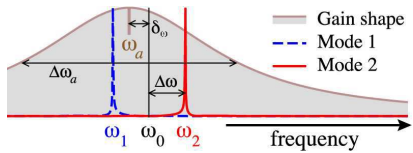}

\caption{(Color online) Schematic illustration of the mode frequencies ($\omega_{1,2}\equiv\omega_{0}\mp\Delta\omega$)
with respect to gain ($\delta_{\omega}=\omega_{a}-\omega_{0}$), as
used throughout the paper.\label{fig:FREQUENCIES}}
\end{figure}
The semiclassical laser equations used in the present paper as a starting
point are composed of three parts: (i) the laser rate equations, reduced
to the equation for population inversion~$W$ of the laser transition;
(ii) the equation of motion for the macroscopic polarization density~$P$
of the laser medium, obtained in a modified electronic oscillator
model, and (iii) the scalar wave equation derived from the Maxwell
equations. We consider two-dimensional (2D) systems, translationally
invariant in the $\hat{\mathbf{z}}$-direction, with TM light polarization,
corresponding to a wide range of 2D photonic structures. In this case
the electric field is $\mathbf{E}(\mathbf{r})=E_{z}(x,z)\hat{\mathbf{z}}$,
allowing us to restrict ourselves to the $z$-component of the field
$E(\mathbf{r},t)=E_{z}(x,y,t)$. Applying the slowly varying envelope
(SVE) approximation \cite{Siegman}, the Maxwell-Bloch system of equations
takes the form \cite{mcHodges}\begin{widetext}\begin{eqnarray}
\frac{\partial}{\partial t}W(\mathbf{r},t) & = & \gamma_{\parallel}\left[R-W(\mathbf{r},t)\right]+\frac{\mathrm{i}}{4\hbar}\left[E(\mathbf{r},t)P^{*}(\mathbf{r},t)-E^{*}(\mathbf{r},t)P(\mathbf{r},t)\right],\label{eq:SVEA_W}\end{eqnarray}
\begin{eqnarray}
\frac{\partial}{\partial t}P(\mathbf{r},t) & = & -\left(\gamma_{\perp}+\mathrm{i}\delta\right)P(\mathbf{r},t)-\frac{\mathrm{i}\mu^{2}}{\hbar}W(\mathbf{r},t)E(\mathbf{r},t),\label{eq:SVEA_P}\end{eqnarray}
\begin{eqnarray}
\frac{1}{\epsilon_{0}}\frac{\partial^{2}}{\partial t^{2}}\left(P(\mathbf{r},t)\mathrm{e}^{-\mathrm{i}\omega t}\right) & = & \left[c^{2}\nabla^{2}-\epsilon(\mathbf{r})\frac{\partial^{2}}{\partial t^{2}}-\kappa(\mathbf{r})\frac{\partial}{\partial t}\right]\left(E(\mathbf{r},t)\mathrm{e}^{-\mathrm{i}\omega t}.\right)\label{eq:SVEA_E}\end{eqnarray}

\end{widetext}Here $W(\mathbf{r},t)$ has the meaning of population
inversion, which can vary spatially as opposed to Ref.~\cite{mcWieczorekPRA}
where it is assumed to be constant across the whole cavity. Further,
$R$~is the external pumping rate, $\mu$~is the dipole matrix element
of the atomic laser transition, and the polarization and population
inversion decay rates are given by $\gamma_{\perp}$~and~$\gamma_{\parallel}$,
respectively. We consider a resonant system that features two eigenmodes
with decay rates $\kappa_{1,2}$, phenomenologically accounted for
by the presence of a loss term $\kappa(\mathbf{r})$ in Eq.~\eqref{eq:SVEA_E}.
The mode frequencies are $\omega_{1,2}\equiv\omega_{0}\mp\Delta\omega$,
and the central frequency~$\omega_{0}$ is shifted with respect to
the lasing transition frequency~$\omega_{a}$ by $\delta_{\omega}=\omega_{a}-\omega_{0}$
with $\Delta\omega,\delta_{\omega}\ll\omega_{0}$, as shown in Fig.~\ref{fig:FREQUENCIES}.
We assume that the eigenmodes of the cold cavity have a spatial structure
given by~$u_{1,2}(\mathbf{r})$. The electric field~$E(\mathbf{r},t)$
is then decomposed into the spatially dependent mode profiles~$u_{1,2}(\mathbf{r})$
multiplied by time dependent SVE functions~$E_{1,2}(t)$ as \begin{equation}
\begin{aligned}E(\mathbf{r},t)\mathrm{e}^{-\mathrm{i}\omega_{0}t}=u_{1}(\mathbf{r})E_{1}(t)\mathrm{e}^{-\mathrm{i}\omega_{1}t}+u_{2}(\mathbf{r})E_{2}(t)\mathrm{e}^{-\mathrm{i}\omega_{2}t}\\
\equiv\left[u_{1}(\mathbf{r})E_{1}(t)\mathrm{e}^{\phi_{+}}+u_{2}(\mathbf{r})E_{2}(t)\mathrm{e}^{\phi_{-}}\right]\mathrm{e}^{-\mathrm{i}\omega_{0}t}.\end{aligned}
\label{eq:decomp_E}\end{equation}
Here and further, $\phi_{\pm}\equiv\pm\mathrm{i}\Delta\omega t$.
Following the approach in \cite{mcHodges}, we make a similar ansatz
for the polarization, introducing the amplitudes~$P_{1,2}(t)$ as
\begin{equation}
P(\mathbf{r},t)\mathrm{e}^{-\mathrm{i}\omega_{0}t}=\left[u_{1}(\mathbf{r})P_{1}(t)\mathrm{e}^{\phi_{+}}+u_{2}(\mathbf{r})P_{2}(t)\mathrm{e}^{\phi_{-}}\right]\mathrm{e}^{-\mathrm{i}\omega_{0}t}.\label{eq:decomp_P}\end{equation}

The applicability of the expansion~\eqref{eq:decomp_P} needs further
justification. Eq.~\eqref{eq:decomp_P} assumes that polarization
$P(\mathbf{r},t)$ and the electric field $E(\mathbf{r},t)$ have
similar spatial profiles. This is strictly true only if the field
intensity is small enough, e.g., if the pumping rate~$R$ is not
very large. Otherwise, the polarization gets influenced by the saturation
terms that involve the population inversion~$W(\mathbf{r},t)$, which
itself %
\begin{comment}
does not distinguish between the modes and thus 
\end{comment}
{}cannot be spatially decomposed. These saturation terms would modify
the spatial profile of $P(\mathbf{r},t)$ outside the scope of Eq.~\eqref{eq:decomp_P}. 

However, as Eq.~\eqref{eq:decomp_P} does not contain any explicit
expansion in a series of nonlinearity orders with subsequent series
truncation, the constraint on the pumping rate~$R$ appears to be
much weaker than what is enforced by the usual near-threshold expansion
\cite{mcZehnle,Haken,HakenFu}, which explicitly retains only third-order
nonlinearities in the hole burning interaction. In the extreme (single-mode)
case, where Eq.~\eqref{eq:decomp_P} implies $P(\mathbf{r},t)\propto E(\mathbf{r},t)$
and thus carries the strongest approximation, it can be shown that
the coupled mode theory based on Eq.~\eqref{eq:decomp_P} leads to
underestimation of the steady-state laser field intensity $E(R)$.
However, the character of the dependence $E(R)$ is preserved for
the values of~$R$ well outside the range of applicability of the
near-threshold expansion (see \cite{Tureci06}). Moreover, the dynamical
behavior of the laser is also correctly predicted by the coupled mode
theory employing the expansion~\eqref{eq:decomp_P} both for one
and for two modes (see our earlier work \cite{zhukFDTD} for a comparison
with direct numerical simulations).

That taken into account, in what follows we will use the expansion
\eqref{eq:decomp_P}, remembering that the results may deviate quantitatively
and may be subjest to further checking as the pumping rate goes far
above threshold.

{}

\subsection{Class-C lasers\label{sub:EQS-C}}

In order to derive the equations for $E_{i}(t)$ and $P_{i}(t)$,
one has to eliminate all the spatial dependencies from Eqs.~\eqref{eq:SVEA_W}--\eqref{eq:SVEA_E}.
We begin by substituting Eqs.~\eqref{eq:decomp_E} and~\eqref{eq:decomp_P}
into Eq.~\eqref{eq:SVEA_E}, assuming that the time dependence of
the field envelopes are slow enough so that $\left|\mathrm{d}E_{j}/\mathrm{d}t\right|\ll\omega_{j}\left|E_{j}\right|$.
The modes~$u_{j}(\mathbf{r})$ are assumed to be orthonormal solutions
of the homogeneous wave equation $(c^{2}\nabla^{2}-\epsilon(\mathbf{r})\omega_{j}^{2})u_{j}(\mathbf{r})=0$,
which means that their integral across the cavity is \begin{equation}
\int_{C}\varepsilon(\mathbf{r})u_{i}^{*}(\mathbf{r})u_{j}(\mathbf{r})=\delta_{ij}.\label{eq:orthogonal_cavity}\end{equation}
As a result, the spatial derivatives in Eq.~\eqref{eq:SVEA_E} can
be eliminated. If $\epsilon(\mathbf{r})=\epsilon$ is constant throughout
the cavity (the bulk-cavity case \cite{mcHodges}), the modes in Eq.~
\eqref{eq:SVEA_E} decouple rigorously, and one obtains%
\begin{comment}
OLD VERSION: Taking into account that the modes~$u_{j}(\mathbf{r})$
are the solutions of the homogeneous wave equation assumed to form
an orthonormal set in the cavity, one can substitute Eqs.~\eqref{eq:decomp_E}
and~\eqref{eq:decomp_P} into Eq.~\eqref{eq:SVEA_E} with subsequent
elimination of the spatial derivatives to disentangle the equations
for each mode~$E_{j}$ under the assumption that all time dependencies
are slow enough. We refer the reader to our earlier work \cite{zhukFDTD}
for details. Here we note that this decoupling is rigorous if $\epsilon$~is
constant throughout the cavity (the bulk-cavity case \cite{mcHodges})
and remains approximately true if the majority of the modes' energy,
as well as the laser active medium, are located in a material with
the same dielectric constant. This is often the case in microcavities.
A more complicated case of distributed feedback structures would require
additional spatial multiscale analysis, e.g., following the approach
developed for photonic crystal lasers \cite{mcJohnBusch}. That kept
in mind, one obtains (compare \cite{mcHodges})
\end{comment}
{}

\begin{equation}
\frac{\mathrm{d}}{\mathrm{d}t}E_{j}=-\frac{\kappa_{j}E_{j}}{2}+\mathrm{\frac{i}{2\epsilon_{0}\epsilon}}\omega_{j}P_{j}.\label{eq:Ec}\end{equation}
This decoupling remains approximately true if the major part of the
modes' energy are located in a material with the same dielectric constant,
as is often the case in microcavities. For details we refer the reader
to our earlier work \cite{zhukFDTD}. A more complicated case of distributed
feedback structures would require additional spatial multiscale analysis,
e.g., following the approach developed for photonic crystal lasers
\cite{mcJohnBusch}.

Eliminating spatial dependencies in Eq.~\eqref{eq:SVEA_P} is simpler
and requires substitution of Eqs.~\eqref{eq:decomp_E}--\eqref{eq:decomp_P}
with subsequent projection onto the eigenmodes, i.e., integration
$\int u_{j}^{*}(\ldots)\mathrm{d}^{3}\mathbf{r}$ over the gain medium:
\begin{equation}
\begin{aligned}\frac{\mathrm{d}}{\mathrm{d}t}P_{1}=\: & -\beta_{1}P_{1}-\mathrm{i}\frac{\mu^{2}}{\hbar}\left(E_{1}W_{11}+E_{2}W_{12}\mathrm{e}^{2\phi_{-}}\right),\\
\frac{\mathrm{d}}{\mathrm{d}t}P_{2}=\: & -\beta_{2}P_{2}-\mathrm{i}\frac{\mu^{2}}{\hbar}\left(E_{1}W_{21}\mathrm{e}^{2\phi_{+}}+E_{2}W_{22}\right),\end{aligned}
\label{eq:Pc}\end{equation}
where $\beta_{1,2}=\left(\gamma_{\perp}+\mathrm{i}\delta_{\omega}\right)\pm\mathrm{i}\Delta\omega$
and $W_{ij}$~are the projections of the population inversion $W(\mathbf{r},t)$
onto the corresponding modes\begin{equation}
W_{ij}(t)\equiv\epsilon\int_{G}\mathrm{d}^{3}\mathbf{r}\, u_{i}^{*}(\mathbf{r})W(\mathbf{r},t)u_{j}(\mathbf{r})\label{eq:W_comps}\end{equation}
Analogously, by substituting Eqs.~\eqref{eq:decomp_E}--\eqref{eq:decomp_P}
into Eq.~\eqref{eq:SVEA_W} and applying $\int u_{i}^{*}(\ldots)u_{j}\mathrm{d}^{3}\mathbf{r}$,
one can obtain the equations for~$W_{ij}$ in the following form:\begin{widetext}\begin{equation}
\begin{aligned}\frac{\mathrm{d}}{\mathrm{d}t}W_{ij}=\: & \gamma_{\parallel}\left(R_{ij}-W_{ij}\right)\\
-\, & \frac{\mathrm{i}}{4\hbar}\left[E_{1}^{*}\left(\alpha_{ij}^{11}P_{1}+\alpha_{ij}^{12}P_{2}\mathrm{e}^{2\phi_{-}}\right)+E_{2}^{*}\left(\alpha_{ij}^{21}P_{1}\mathrm{e}^{2\phi_{+}}+\alpha_{ij}^{22}P_{2}\right)\right]\\
+\, & \frac{\mathrm{i}}{4\hbar}\left[E_{1}\left(\alpha_{ij}^{11}P_{1}^{*}+\alpha_{ij}^{21}P_{2}^{*}\mathrm{e}^{2\phi_{+}}\right)+E_{2}\left(\alpha_{ij}^{12}P_{1}^{*}\mathrm{e}^{2\phi_{-}}+\alpha_{ij}^{22}P_{2}^{*}\right)\right].\end{aligned}
\label{eq:Wc}\end{equation}
Here, $R_{ij}$ are related to $R$ in the same way as $W_{ij}$ to
$W(\mathbf{r},t)$, via Eq.~\eqref{eq:W_comps}. The coefficients
$\alpha_{ij}^{mn}$ are mode overlap integrals defined as:\begin{equation}
\alpha_{ij}^{mn}\equiv\epsilon\int_{G}\mathrm{d}^{3}\mathbf{r}\, u_{i}^{*}(\mathbf{r})u_{j}(\mathbf{r})u_{m}^{*}(\mathbf{r})u_{n}(\mathbf{r}).\label{eq:alpha_comps}\end{equation}
The integration in Eqs.~\eqref{eq:W_comps} and~\eqref{eq:alpha_comps}
is performed over the gain medium where $\epsilon(\mathbf{r})=\epsilon$~is
assumed to be constant. Apart from that assumption, the shape of the
gain region itself can be arbitrary and does not have to be contiguous.
The mode geometry can also be arbitrary unlike in the previous reports
\cite{mcHodges,mcWieczorekOC,mcWieczorekPRA}, as the inter-mode and
mode-gain overlaps are accounted for in terms of~$\alpha_{ij}^{mn}$
and~$W_{ij}$. Note that Eqs.~\eqref{eq:Pc} and~\eqref{eq:Wc}
with the definition~\eqref{eq:W_comps} do not involve any approximations
on the field or pump intensity beside the one associated with the
validity of Eq.~\eqref{eq:decomp_P} as described above. Because
of this, the full population inversion $W(\mathbf{r},t)$ cannot be
written explicitly in terms of $W_{ij}(t)$ and $u_{1,2}(\mathbf{r})$,
in contrast to $E$~and~$P$, as in Eqs.~\eqref{eq:decomp_E}--\eqref{eq:decomp_P}.
Also note that the rate equations~\eqref{eq:Wc} for the population
inversion explicitly contain oscillatory terms, which originate from
the beating in the superposition of the two modes with different frequencies
$\omega_{1}$~and~$\omega_{2}$.

\subsection{Class-B lasers\label{sub:EQS-B}}

Equations \eqref{eq:Ec},~\eqref{eq:Pc}, and~\eqref{eq:Wc} govern
the dynamics of the two spectrally close, interacting modes without
any assumptions on the laser dynamics besides those needed for the
SVE approximation. All these equations include a decay term with a
characteristic decay rate for all the variables involved. The mode
amplitudes~$E_{j}$ decay with the rate~$\kappa_{j}$ associated
with the $Q$-factors of the modes ($Q_{j}=\omega_{j}/\kappa_{j}$).
The decay of all the population inversion projections~$W_{ij}$ is
governed by~$\gamma_{\parallel}$. Finally, the polarization amplitudes~$P_{j}$
decay rates are complex, $\beta_{j}=\left(\gamma_{\perp}+\mathrm{i}\delta_{\omega}\right)\pm\mathrm{i}\Delta\omega$.
This complexity directly results from the multimode character of the
laser under study, and in the single-mode case $\beta_{j}=\gamma_{\perp}$.

In the most general case of laser dynamics there are no restrictions
on the decay rates $\kappa_{j}$, $\gamma_{\parallel}$. $\gamma_{\perp}$
(so-called class-C lasers). In reality, however, the decay rates are
governed by different physical processes and often belong to different
time scales (class-B or class-A lasers, see \cite{mcWieczorekOC}),
which can make the analysis of the laser equations considerably simpler.

Class-B lasers are defined by $\gamma_{\perp}\gg\gamma_{\parallel},\kappa_{j}$.
In the single-mode case, it would mean that the polarization relaxes
and achieves saturation so fast that the polarization can be assumed
to have no own dynamics and follows $E$~and~$W$ adiabatically.

In the two-mode case, where the polarization dynamics is influenced
by the intermode spacing~$\Delta\omega$, the introduction of the
class-B approximations needs to be approached with greater care. Since
the right-hand side of Eqs.~\eqref{eq:Pc} includes oscillatory terms
on the time scale of $2\Delta\omega$, one can eliminate the polarization
only if these oscillations are much slower than the exponential decay
due to~$\gamma_{\perp}$, i.e., $\gamma_{\perp}\gg\Delta\omega$.
Note that this additional condition for class-B lasing, specific for
multimode lasers, becomes especially important in microlasers where
the small cavity size can place the modes much further apart from
each other than in the bulk cavities. 

Under these assumptions, we can now eliminate the polarization adiabatically
by assuming $\mathrm{d}P_{j}/\mathrm{d}t\approx0$. Hence, Eqs.~\eqref{eq:Pc}
assume the form\begin{equation}
\begin{aligned}P_{1}=\: & -\mathrm{i}\frac{\mu^{2}}{\hbar}\frac{1}{\beta_{1}}\left(E_{1}W_{11}+E_{2}W_{12}\mathrm{e}^{2\phi_{-}}\right),\\
P_{2}=\: & -\mathrm{i}\frac{\mu^{2}}{\hbar}\frac{1}{\beta_{2}}\left(E_{1}W_{21}\mathrm{e}^{2\phi_{+}}+E_{2}W_{22}\right),\end{aligned}
\label{eq:Pb}\end{equation}
which causes Eqs.~\eqref{eq:Ec} to be modified as\begin{equation}
\begin{aligned}\frac{\mathrm{d}}{\mathrm{d}t}E_{1}=\: & -\frac{\kappa_{1}}{2}E_{1}+\mathrm{\frac{\mu^{2}}{\hbar}\frac{\omega_{1}}{2\epsilon_{0}\epsilon}\frac{1}{\beta_{1}}}\left(E_{1}W_{11}+E_{2}W_{12}\mathrm{e}^{2\phi_{-}}\right),\\
\frac{\mathrm{d}}{\mathrm{d}t}E_{2}=\: & -\frac{\kappa_{2}}{2}E_{2}+\frac{\mu^{2}}{\hbar}\frac{\omega_{2}}{2\epsilon_{0}\epsilon}\frac{1}{\beta_{2}}\left(E_{1}W_{21}\mathrm{e}^{2\phi_{+}}+E_{2}W_{22}\right).\end{aligned}
\label{eq:Eb}\end{equation}
Analogously, substituting Eq.~\eqref{eq:Pb} into Eq.~\eqref{eq:Wc}
one may obtain the equations for~$W_{ij}$. Since the population
inversion~$W$ is real {[}see Eq.~\eqref{eq:SVEA_W}], it follows
from Eq.~\eqref{eq:W_comps} that $W_{ji}^{*}=W_{ij}$, and in particular,
$W_{jj}^{*}=W_{jj}$. Hence, \begin{equation}
\begin{aligned}\frac{\mathrm{d}}{\mathrm{d}t}W_{ij}=\: & \gamma_{\parallel}\left(R_{ij}-W_{ij}\right)\\
-\, & \frac{\mu^{2}}{4\hbar^{2}}\left|E_{1}\right|^{2}\left[\alpha_{ij}^{11}\left(\frac{1}{\beta_{1}}+\frac{1}{\beta_{1}^{*}}\right)W_{11}+\left(\frac{\alpha_{ij}^{12}}{\beta_{2}}W_{21}+\frac{\alpha_{ij}^{21}}{\beta_{2}^{*}}W_{12}\right)\right]\\
-\, & \frac{\mu^{2}}{4\hbar^{2}}\left|E_{2}\right|^{2}\left[\alpha_{ij}^{22}\left(\frac{1}{\beta_{2}}+\frac{1}{\beta_{2}^{*}}\right)W_{22}+\left(\frac{\alpha_{ij}^{21}}{\beta_{1}}W_{12}+\frac{\alpha_{ij}^{12}}{\beta_{1}^{*}}W_{21}\right)\right]\\
-\, & \frac{\mu^{2}}{4\hbar^{2}}E_{1}^{*}E_{2}\left[\left(\frac{\alpha_{ij}^{11}}{\beta_{1}}+\frac{\alpha_{ij}^{22}}{\beta_{2}^{*}}\right)W_{12}+\alpha_{ij}^{12}\left(\frac{1}{\beta_{2}}W_{22}+\frac{1}{\beta_{1}^{*}}W_{11}\right)\right]\mathrm{e}^{2\phi_{-}}\\
-\, & \frac{\mu^{2}}{4\hbar^{2}}E_{1}E_{2}^{*}\left[\left(\frac{\alpha_{ij}^{22}}{\beta_{2}}+\frac{\alpha_{ij}^{11}}{\beta_{1}^{*}}\right)W_{21}+\alpha_{ij}^{21}\left(\frac{1}{\beta_{1}}W_{11}+\frac{1}{\beta_{2}^{*}}W_{22}\right)\right]\mathrm{e}^{2\phi_{+}}.\end{aligned}
\label{eq:Wb}\end{equation}

Eqs.~\eqref{eq:Eb}--\eqref{eq:Wb} are the governing equations for
two-mode class-B lasers. Further knowledge about the modes in question
can allow further simplification. A good example is the case when
the modes are orthogonal not only withinin the whole cavity {[}Eq.~\eqref{eq:orthogonal_cavity}],
but also in the gain region, e.g., if most of the cavity or at least
the portion of the cavity with maximum mode energy is filled with
the pumped gain medium: \begin{eqnarray}
\int_{G}u_{i}^{*}(\mathbf{r})u_{j}(\mathbf{r}) & = & \delta_{ij}.\label{eq:orthogonal_gain}\end{eqnarray}
In this case the overlap integrals with one out-of-place index ($\alpha_{ij}^{ii}$,~$\alpha_{ii}^{ji}$,~etc.)
will be negligible compared to the rest of the overlaps such as $\alpha_{jj}^{jj}$,
$\alpha_{jj}^{ii}$, $\alpha_{ij}^{ij}$, or $\alpha_{ji}^{ij}$.
This allows to shorten Eq.~\eqref{eq:Wb}, which then assume different
forms for symmetric~$W_{jj}$ vs. anti-symmetric projections~$W_{ij\neq i}$:\begin{equation}
\begin{aligned}\frac{\mathrm{d}}{\mathrm{d}t}W_{jj}=\: & \gamma_{\parallel}\left(R_{jj}-W_{jj}\right)-\frac{\mu^{2}}{4\hbar^{2}}\left[\left|E_{1}\right|^{2}\alpha_{jj}^{11}\left(\frac{1}{\beta_{1}}+\frac{1}{\beta_{1}^{*}}\right)W_{11}+\left|E_{2}\right|^{2}\alpha_{jj}^{22}\left(\frac{1}{\beta_{2}}+\frac{1}{\beta_{2}^{*}}\right)W_{22}\right]\\
-\, & \frac{\mu^{2}}{4\hbar^{2}}\left[E_{1}^{*}E_{2}\mathrm{e}^{2\phi_{-}}\left(\frac{\alpha_{jj}^{11}}{\beta_{1}}+\frac{\alpha_{jj}^{22}}{\beta_{2}^{*}}\right)W_{12}+E_{1}E_{2}^{*}\mathrm{e}^{2\phi_{+}}\left(\frac{\alpha_{jj}^{11}}{\beta_{1}^{*}}+\frac{\alpha_{jj}^{22}}{\beta_{2}}\right)W_{21}\right],\end{aligned}
\label{eq:Wb_jj}\end{equation}
\begin{equation}
\begin{aligned}\frac{\mathrm{d}}{\mathrm{d}t}W_{12}=\: & \gamma_{\parallel}\left(R_{12}-W_{12}\right)-\frac{\mu^{2}}{4\hbar^{2}}\left[\left|E_{1}\right|^{2}\left(\frac{\alpha_{12}^{12}}{\beta_{2}}W_{21}+\frac{\alpha_{12}^{21}}{\beta_{2}^{*}}W_{12}\right)+\left|E_{2}\right|^{2}\left(\frac{\alpha_{12}^{21}}{\beta_{1}}W_{12}+\frac{\alpha_{12}^{12}}{\beta_{1}^{*}}W_{21}\right)\right]\\
-\, & \frac{\mu^{2}}{4\hbar^{2}}\left[E_{1}^{*}E_{2}\mathrm{e}^{2\phi_{-}}\alpha_{12}^{12}\left(\frac{1}{\beta_{2}}W_{22}+\frac{1}{\beta_{1}^{*}}W_{11}\right)+E_{1}E_{2}^{*}\mathrm{e}^{2\phi_{+}}\alpha_{12}^{21}\left(\frac{1}{\beta_{2}^{*}}W_{22}+\frac{1}{\beta_{1}}W_{11}\right)\right].\end{aligned}
\label{eq:Wb_12}\end{equation}
where $R_{12}\ll R_{jj}$ due to the mode orthogonality and, as we
remember, $W_{21}=W_{12}^{*}$. Furthermore, if the modes are \emph{intensity-matched},
i.e., assumed to have nearly equal intensity distribution in the gain
region so that\begin{equation}
\left|u_{1}(\mathbf{r})\right|^{2}\approx\left|u_{2}(\mathbf{r})\right|^{2},\quad\mathbf{r}\in G,\label{eq:intensities}\end{equation}
then it follows from Eq.~\eqref{eq:W_comps} that $W_{11}=W_{22}\equiv W_{s}$
and $W_{12}=W_{21}^{*}\equiv W_{a}$, as well as from Eq.~\eqref{eq:alpha_comps}
that $\alpha_{jj}^{ii}=\alpha_{ji}^{ij}\equiv\alpha$ is real, while
$\alpha_{ij}^{ij}\equiv\alpha'$ can be complex. Hence,\begin{equation}
\begin{aligned}\frac{\mathrm{d}}{\mathrm{d}t}W_{s}=\: & \gamma_{\parallel}\left(R_{s}-W_{s}\right)-\frac{\mu^{2}}{4\hbar^{2}}\alpha\left[\left|E_{1}\right|^{2}\left(\frac{1}{\beta_{1}}+\frac{1}{\beta_{1}^{*}}\right)+\left|E_{2}\right|^{2}\left(\frac{1}{\beta_{2}}+\frac{1}{\beta_{2}^{*}}\right)\right]W_{s}\\
-\, & \frac{\mu^{2}}{4\hbar^{2}}\alpha\left[E_{1}^{*}E_{2}\mathrm{e}^{2\phi_{-}}\left(\frac{1}{\beta_{1}}+\frac{1}{\beta_{2}^{*}}\right)W_{a}+E_{1}E_{2}^{*}\mathrm{e}^{2\phi_{+}}\left(\frac{1}{\beta_{1}^{*}}+\frac{1}{\beta_{2}}\right)W_{a}^{*}\right],\end{aligned}
\label{eq:Wb_s}\end{equation}
\begin{equation}
\begin{aligned}\frac{\mathrm{d}}{\mathrm{d}t}W_{a}=\: & -\gamma_{\parallel}W_{a}-\frac{\mu^{2}}{4\hbar^{2}}\left[E_{1}^{*}E_{2}\mathrm{e}^{2\phi_{-}}\alpha'\left(\frac{1}{\beta_{2}}+\frac{1}{\beta_{1}^{*}}\right)+E_{1}E_{2}^{*}\mathrm{e}^{2\phi_{+}}\alpha\left(\frac{1}{\beta_{2}^{*}}+\frac{1}{\beta_{1}}\right)\right]W_{s}\\
-\, & \frac{\mu^{2}}{4\hbar^{2}}\left[\left|E_{1}\right|^{2}\left(\frac{\alpha'}{\beta_{2}}W_{a}^{*}+\frac{\alpha}{\beta_{2}^{*}}W_{a}\right)+\left|E_{2}\right|^{2}\left(\frac{\alpha}{\beta_{1}}W_{a}+\frac{\alpha'}{\beta_{1}^{*}}W_{a}^{*}\right)\right].\end{aligned}
\label{eq:Wb_a}\end{equation}

%
\begin{comment}
OLD: One can see that the right-hand sides of the equation for~$W_{s}$
\eqref{eq:Wb_s} contains~$W_{s}$ with a coefficient which is {}``incoherent''
with respect to the modes, i.e., involving only the mode intensities
$\left|E_{j}\right|^{2}$, while~$W_{a}$ is involved in a more complicated,
{}``coherent'' fashion, involving intermode phase (contained in
the products $E_{i}^{*}E_{j}$) as well as explicit oscillatory exponents
$\mathrm{e}^{2\phi_{\pm}}$. Correspondingly, the equation for~$W_{a}$
\eqref{eq:Wb_a} contains~$W_{a}$ in a {}``coherent'' and~$W_{s}$
in an {}``incoherent'' fashion.
\end{comment}
{}

\subsection{Class-A lasers\label{sub:EQS-A}}

If one further assumes that $(\gamma_{\perp}\gg)\gamma_{\parallel}\gg\kappa_{j}$
(class-A lasers), the slowest-varying quantity becomes the mode decay.
The population inversion follows the mode amplitudes~$E_{j}(t)$
instantaneously and can be eliminated, leaving us with only two equations
for the mode amplitudes. Similar to the way we have built the class-B
approximation, the derivatives in Eqs.~\eqref{eq:Wb} are $\mathrm{d}W_{ij}/\mathrm{d}t\approx0$.
In this case, Eqs.~\eqref{eq:Wb_jj}--\eqref{eq:Wb_12} become\begin{equation}
\begin{aligned}W_{jj}=\: & R_{jj}-\frac{\mu^{2}}{4\hbar^{2}}\frac{1}{\gamma_{\parallel}}\left[\left|E_{1}\right|^{2}\alpha_{jj}^{11}\left(\frac{1}{\beta_{1}}+\frac{1}{\beta_{1}^{*}}\right)W_{11}+\left|E_{2}\right|^{2}\alpha_{jj}^{22}\left(\frac{1}{\beta_{2}}+\frac{1}{\beta_{2}^{*}}\right)W_{22}\right]\\
-\, & \frac{\mu^{2}}{4\hbar^{2}}\left[E_{1}^{*}E_{2}\mathrm{e}^{2\phi_{-}}\left(\frac{\alpha_{jj}^{11}}{\beta_{1}}+\frac{\alpha_{jj}^{22}}{\beta_{2}^{*}}\right)W_{12}+E_{1}E_{2}^{*}\mathrm{e}^{2\phi_{+}}\left(\frac{\alpha_{jj}^{11}}{\beta_{1}^{*}}+\frac{\alpha_{jj}^{22}}{\beta_{2}}\right)W_{21}\right],\end{aligned}
\label{eq:Wa_jj}\end{equation}
\begin{equation}
\begin{aligned}W_{12}=\: & -\frac{\mu^{2}}{4\hbar^{2}}\frac{1}{\gamma_{\parallel}}\left[\left|E_{1}\right|^{2}\left(\frac{\alpha_{12}^{12}}{\beta_{2}}W_{21}+\frac{\alpha_{12}^{21}}{\beta_{2}^{*}}W_{12}\right)+\left|E_{2}\right|^{2}\left(\frac{\alpha_{12}^{21}}{\beta_{1}}W_{12}+\frac{\alpha_{12}^{12}}{\beta_{1}^{*}}W_{21}\right)\right]\\
-\, & \frac{\mu^{2}}{4\hbar^{2}}\left[E_{1}^{*}E_{2}\mathrm{e}^{2\phi_{-}}\alpha_{12}^{12}\left(\frac{1}{\beta_{2}}W_{22}+\frac{1}{\beta_{1}^{*}}W_{11}\right)+E_{1}E_{2}^{*}\mathrm{e}^{2\phi_{+}}\alpha_{12}^{21}\left(\frac{1}{\beta_{2}^{*}}W_{22}+\frac{1}{\beta_{1}}W_{11}\right)\right].\end{aligned}
\label{eq:Wa_12}\end{equation}
This is a system of linear algebraic equations that can be solved
for~$W_{ij}$. We are aiming for equations with simple enough structure
to be treated analytically, namely, equations for~$E_{j}$ with up
to cubic-order non-linearity as analyzed, e.g., in \cite{Siegman}.
Hence, we are looking for the solutions in the form \begin{equation}
W_{ij}\equiv W_{ij}^{(0)}+\sum_{m,n}W_{ij}^{(m,n)}E_{m}^{*}E_{n},\label{eq:Wa_ansatz}\end{equation}
neglecting terms with higher powers of~$E$. Truncating higher-order
nonlinearity corresponds physically to the case with low field intensities,
i.e., just above the lasing threshold. Hence, at this point the near-threshold
expansion is introduced as understood in numerous works \cite{mcZehnle,Haken,HakenFu}.
We remark that this expansion is by far a stronger approximation than
the one used in assuming the form~\eqref{eq:decomp_P} for the polarization.
Hence, the class-B and class-C models described in the previous sections
are valid for much stronger pumping, while the class-A description
that follows is valid for pumping rates only slightly above threshold.
Inserted into Eqs.~\eqref{eq:Wa_jj}--\eqref{eq:Wa_12}, Eq.~\eqref{eq:Wa_ansatz}
yields\begin{equation}
\begin{aligned}W_{jj}\approx\: & R_{jj}-\frac{\mu^{2}}{4\hbar^{2}}\frac{1}{\gamma_{\parallel}}\left[\left|E_{1}\right|^{2}\alpha_{jj}^{11}\left(\frac{1}{\beta_{1}}+\frac{1}{\beta_{1}^{*}}\right)R_{11}+\left|E_{2}\right|^{2}\alpha_{jj}^{22}\left(\frac{1}{\beta_{2}}+\frac{1}{\beta_{2}^{*}}\right)R_{22}\right],\\
W_{12}\approx\: & -\frac{\mu^{2}}{4\hbar^{2}}\frac{1}{\gamma_{\parallel}}\left[E_{1}^{*}E_{2}\mathrm{e}^{2\phi_{-}}\alpha_{12}^{12}\left(\frac{1}{\beta_{2}}R_{22}+\frac{1}{\beta_{1}^{*}}R_{11}\right)+E_{1}E_{2}^{*}\mathrm{e}^{2\phi_{+}}\alpha_{12}^{21}\left(\frac{1}{\beta_{2}^{*}}R_{22}+\frac{1}{\beta_{1}}R_{11}\right)\right].\end{aligned}
\label{eq:eqs_Wa_approx}\end{equation}
Note that the right-hand side of Eqs.~\eqref{eq:Wa_jj}--\eqref{eq:Wa_12}
has terms of the form $E_{m}^{*}E_{n}W_{ij}$. Hence the same result
could be obtained by solving the equation system $W_{ij}=\mathbb{L}\cdot W_{ij}$
iteratively as $W_{ij}^{(k)}=\mathbb{L}\cdot W_{ij}^{(k-1)}$ with
$W_{ij}^{(0)}=0$ up to~$W_{ij}^{(2)}$, as was done in \cite{mcHodges,mcWieczorekPRA,zhukPRL}.

Note that the presence of oscillatory exponents~$\mathrm{e}^{2\phi_{\pm}}$
on the right-hand side of Eqs.~\eqref{eq:Wb}, induced by beating
of the field intensities, dictates that an adiabatic elimination can
only be performed safely if $\gamma_{\parallel}\gg\Delta\omega$.
Unfortunately, this assumption is quite restrictive and makes the
resulting class-A laser equations hardly applicable for any two-mode
system beyond the case of spectrally overlapping modes unless the
mode $Q$-factors become very high. However, Eq.~\eqref{eq:eqs_Wa_approx}
suggests that $W_{12}$~should be oscillatory with frequency~$2\Delta\omega$.
This is indeed the case, as confirmed by numerical solution of class-B
or class-C equations. These oscillations (also called population pulsations)
are the main reason why the condition $\mathrm{d}W_{12}/\mathrm{d}t\approx0$
is valid only for vanishingly small~$\Delta\omega$. By accounting
for these pulsations explicitly, one can build class-A laser equations
applicable for a wider range of~$\Delta\omega$. We introduce oscillatory
terms $\mathrm{e}^{\pm2\mathrm{i}\Delta\omega t}$ into~$W_{12}$:\begin{equation}
W_{12}(t)=W_{21}(t)=\tilde{W}_{a}(t)\mathrm{e}^{2\phi_{+}}+\tilde{W}_{a}^{*}(t)\mathrm{e}^{2\phi_{-}}\label{eq:Wa_12_to_Wa_a}\end{equation}
where the envelope function~$\tilde{W}_{a}(t)$ supposedly varies
more slowly than~$2\Delta\omega$ and on the same time scale as~$W_{jj}(t)$.
We can then reformulate the condition for adiabatic elimination of
$W_{12}$ in the form $\mathrm{d}\tilde{W}_{a}/\mathrm{d}t\approx0$.
The algebraic equation for~$\tilde{W}_{a}$ analogous to \eqref{eq:Wa_12}
is then\begin{equation}
\begin{aligned}\tilde{W}_{a}=\: & -\frac{\mu^{2}}{4\hbar^{2}}\frac{1}{\gamma_{\parallel}+2\mathrm{i}\Delta\omega}\left[\left|E_{1}\right|^{2}\left(\frac{\alpha_{12}^{12}}{\beta_{2}}+\frac{\alpha_{12}^{21}}{\beta_{2}^{*}}\right)\tilde{W}_{a}+\left|E_{2}\right|^{2}\left(\frac{\alpha_{12}^{21}}{\beta_{1}}+\frac{\alpha_{12}^{12}}{\beta_{1}^{*}}\right)\tilde{W}_{a}+E_{1}E_{2}^{*}\alpha_{12}^{21}\left(\frac{1}{\beta_{2}^{*}}W_{22}+\frac{1}{\beta_{1}}W_{11}\right)\right],\end{aligned}
\label{eq:Wa_a}\end{equation}
Note that unlike~$W_{12}$, $\tilde{W}_{a}$~is explicitly complex
due to the substitution $\gamma_{\parallel}\to\gamma_{\parallel}+2\mathrm{i}\Delta\omega$.
Also note the disappearance of oscillatory exponents in Eq.~\eqref{eq:Wa_a},
compared to Eq.~\eqref{eq:Wa_12}. Inserting Eq.~\eqref{eq:Wa_12_to_Wa_a}--\eqref{eq:Wa_a}
into~\eqref{eq:Wa_12} and following the same near-threshold expansion
as above, we obtain the final class-A equations

\begin{equation}
\begin{aligned}\frac{\mathrm{d}}{\mathrm{d}t}E_{1}\approx\: & \left(\frac{g\omega_{1}}{\beta_{1}}R_{1}-\frac{\kappa_{1}}{2}\right)E_{1}-\frac{g\xi\omega_{1}}{\gamma_{\parallel}}\frac{1}{\beta_{1}}\left[\alpha_{11}R_{1}\mathcal{L}_{11}\left|E_{1}\right|^{2}+\alpha_{12}R_{2}\mathcal{L}_{22}\left|E_{2}\right|^{2}\right]E_{1}\\
 & -\frac{g\xi\omega_{1}}{\gamma_{\parallel}+2\mathrm{i}\Delta\omega}\frac{\alpha_{12}}{\beta_{1}}\left(\frac{R_{1}}{\beta_{1}}+\frac{R_{2}}{\beta_{2}^{*}}\right)\left|E_{2}\right|^{2}E_{1}-\frac{g\xi\omega_{1}}{\gamma_{\parallel}-2\mathrm{i}\Delta\omega}\frac{\alpha_{12}}{\beta_{1}}\left(\frac{R_{1}}{\beta_{1}^{*}}+\frac{R_{2}}{\beta_{2}}\right)\left(E_{2}\right)^{2}E_{1}^{*}\mathrm{e}^{4\phi_{-}},\\
\frac{\mathrm{d}}{\mathrm{d}t}E_{2}\approx\: & \left(\frac{g\omega_{2}}{\beta_{2}}R_{2}-\frac{\kappa_{2}}{2}\right)E_{2}-\frac{g\xi\omega_{2}}{\gamma_{\parallel}}\frac{1}{\beta_{2}}\left[\alpha_{22}R_{2}\mathcal{L}_{22}\left|E_{2}\right|^{2}+\alpha_{12}R_{1}\mathcal{L}_{11}\left|E_{1}\right|^{2}\right]E_{2}\\
 & -\frac{g\xi\omega_{2}}{\gamma_{\parallel}-2\mathrm{i}\Delta\omega}\frac{\alpha_{12}}{\beta_{2}}\left(\frac{R_{1}}{\beta_{1}^{*}}+\frac{R_{2}}{\beta_{2}}\right)\left|E_{1}\right|^{2}E_{2}-\frac{g\xi\omega_{2}}{\gamma_{\parallel}+2\mathrm{i}\Delta\omega}\frac{\alpha_{12}}{\beta_{2}}\left(\frac{R_{1}}{\beta_{1}}+\frac{R_{2}}{\beta_{2}^{*}}\right)\left(E_{1}\right)^{2}E_{2}^{*}\mathrm{e}^{4\phi_{+}}.\end{aligned}
\label{eq:Ea_relaxed}\end{equation}
where $g\equiv\mu^{2}/2\epsilon_{0}\epsilon\hbar$, $\xi\equiv\mu^{2}/4\hbar^{2}$,
$\alpha_{jj}\equiv\alpha_{jj}^{jj}$, $\alpha_{12}\equiv\alpha_{jj}^{ii}=\alpha_{ij}^{ji}\approx\alpha_{ij}^{ij}$,
and $\mathcal{L}_{ij}\equiv\beta_{i}^{-1}+\left(\beta_{j}^{*}\right)^{-1}$.
Eqs.~\eqref{eq:Ea_relaxed} retain their applicability for a wide
range of $\Delta\omega$ up to $\Delta\omega\simeq\gamma_{\parallel}$
and beyond. The only limitation is the requirement $\gamma_{\perp}\gg\Delta\omega$
needed to obtain the class-B equations. As was the case with the class-C
to class-B transition, we see that the multimode case needs to be
approached with care, since $\Delta\omega$~represents an additional
dynamical parameter (mode beating). It can play a significant part
in laser dynamics and render some approximations invalid despite their
validity in the single-mode case for the same parameters.

\section{Bistability in class-A microlasers\label{sec:CLASS-A}}

\subsection{Mode competition equations\label{sub:CLA-GEN}}

Now that the dynamics of a two-mode laser has been reduced to relatively
simple class-A equations \eqref{eq:Ea_relaxed}, the mode dynamics
can be analyzed for possible steady-state and stable solutions. Eqs.~\eqref{eq:Ea_relaxed}
resemble the standard 2-mode competition equations (see \cite{Siegman}):
\begin{equation}
\begin{aligned}\frac{\mathrm{d}}{\mathrm{d}t}E_{1}=\left(\rho_{1}-\theta_{11}\left|E_{1}\right|^{2}-\theta_{12}\left|E_{2}\right|^{2}\right)E_{1}-\theta'_{12}\left(E_{2}\right)^{2}E_{1}^{*}\mathrm{e}^{4\phi_{-}},\\
\frac{\mathrm{d}}{\mathrm{d}t}E_{2}=\left(\rho_{2}-\theta_{21}\left|E_{1}\right|^{2}-\theta_{22}\left|E_{2}\right|^{2}\right)E_{2}-\theta'_{21}\left(E_{1}\right)^{2}E_{2}^{*}\mathrm{e}^{4\phi_{+}}.\end{aligned}
\label{eq:2mode_competition}\end{equation}
Here, $\rho_{j}$~in the linear terms characterize the net unsaturated
gain (minus cavity losses) for the mode~$j$. The coefficients $\theta_{jj}$
and $\theta_{ij\neq i}$ are self- and cross-saturation coefficients,
respectively. These terms are fully similar in form and meaning to
the widely studied case in \cite{Siegman}. The last terms, which
are special to Eqs.~\eqref{eq:2mode_competition}, also contribute
to cross-saturation but contain the phases of the modes, as well as
an explicit oscillatory time dependence with frequency~$4\Delta\omega$.
The expressions for all the coefficients can be obtained directly
from Eqs.~\eqref{eq:Ea_relaxed}. 

Since Eqs.~\eqref{eq:2mode_competition} include the phase of the
modes explicitly, they can be separated into amplitude and phase equations.
Substituting $E_{j}(t)=\left|E_{j}(t)\right|\mathrm{e}^{\mathrm{i}\varphi_{j}(t)}$
, one obtains:\begin{equation}
\begin{aligned}\frac{\mathrm{d}}{\mathrm{d}t}\left|E_{1}\right|=\left(\textrm{Re\,}\rho_{1}-\textrm{Re\,}\theta_{11}\left|E_{1}\right|^{2}-\textrm{Re\,}\theta_{12}\left|E_{2}\right|^{2}\right)\left|E_{1}\right|-\textrm{Re}\left(\theta'_{12}\mathrm{e}^{2\mathrm{i}(\varphi_{2}-\varphi_{1})}\mathrm{e}^{4\phi_{-}}\right)\left|E_{2}\right|^{2}\left|E_{1}\right|,\\
\frac{\mathrm{d}}{\mathrm{d}t}\left|E_{2}\right|=\left(\textrm{Re\,}\rho_{2}-\textrm{Re\,}\theta_{21}\left|E_{1}\right|^{2}-\textrm{Re\,}\theta_{22}\left|E_{2}\right|^{2}\right)\left|E_{2}\right|-\textrm{Re}\left(\theta'_{21}\mathrm{e}^{-2\mathrm{i}(\varphi_{2}-\varphi_{1})}\mathrm{e}^{4\phi_{+}}\right)\left|E_{1}\right|^{2}\left|E_{2}\right|,\end{aligned}
\label{eq:2mode_amplitudes}\end{equation}
\begin{equation}
\begin{aligned}\frac{\mathrm{d}}{\mathrm{d}t}\varphi_{1}=\left(\textrm{Im\,}\rho_{1}-\textrm{Im\,}\theta_{11}\left|E_{1}\right|^{2}-\textrm{Im\,}\theta_{12}\left|E_{2}\right|^{2}\right)-\textrm{Im}\left(\theta'_{12}\mathrm{e}^{2\mathrm{i}(\varphi_{2}-\varphi_{1})}\mathrm{e}^{4\phi_{-}}\right)\left|E_{2}\right|^{2},\\
\frac{\mathrm{d}}{\mathrm{d}t}\varphi_{2}=\left(\textrm{Im\,}\rho_{2}-\textrm{Im\,}\theta_{21}\left|E_{1}\right|^{2}-\textrm{Im\,}\theta_{22}\left|E_{2}\right|^{2}\right)-\textrm{Im}\left(\theta'_{21}\mathrm{e}^{-2\mathrm{i}(\varphi_{2}-\varphi_{1})}\mathrm{e}^{4\phi_{+}}\right)\left|E_{1}\right|^{2}.\end{aligned}
\label{eq:2mode_phases}\end{equation}

The amplitude equations \eqref{eq:2mode_amplitudes} now completely
coincide in form with the usual two-mode competition \cite{Siegman}
but contain the intermode phase difference $\Delta\varphi=\varphi_{2}-\varphi_{1}$
as a parameter and have the cross-saturation coefficients explicitly
time-dependent. We can see that the amplitudes always achieve saturation
due to a cubic non-linearity. The phase difference, however, may either
become stationary, corresponding to phase-locked solutions, or be
allowed to vary, in which case the solutions are said to be unlocked.

In the limiting case of $\Delta\omega=0$ one can show that there
are two phase-locked solutions: one stable with $\Delta\varphi=\pi/2$
and one unstable with $\Delta\varphi=0$. Without further assumptions
as to the nature of the modes (such as those in some earlier works
\cite{mcZehnle,mcWieczorekPRA}), the general case is difficult to
analyze due to explicit time dependence in the coefficients for non-zero~$\Delta\omega$.
In particular, $\Delta\omega>0$ causes~$\Delta\varphi$ to undergo
precession even in the locked regimes. As this precession becomes
faster, one can no longer distinguish between locked and unlocked
solutions. For sufficiently large~$\Delta\omega$, the oscillations~$\mathrm{e}^{\pm4\mathrm{i}\Delta\omega t}$
occur fast enough compared to the onset time scale, which primarily
depends on~$\kappa$ rather than on~$\Delta\omega$. In this case
the modes appear always unlocked (mentioned in \cite{mcZehnle} as
a {}``natural tendency'' for different-frequency modes), and the
effects of the phase terms can be averaged out. Our numerical estimations
show that this is possible if $\Delta\omega>10^{-2}\kappa$. The case
$\Delta\omega\ll\kappa$, corresponding to spectrally overlapping
modes, is outside the scope of the present paper anyway as there can
be additional channels of mode coupling (e.g., the Petermann excess
noise \cite{Peterman}). Thus, we will henceforth ignore the phase
terms in Eqs.~\eqref{eq:2mode_competition}--\eqref{eq:2mode_phases}
and rewrite Eq.~\eqref{eq:Ea_relaxed} as \begin{equation}
\begin{aligned}\frac{\mathrm{d}}{\mathrm{d}t}\left|E_{1}\right|\approx\: & \textrm{Re}\left(\frac{g\omega_{1}}{\beta_{1}}R_{1}-\frac{\kappa_{1}}{2}\right)\left|E_{1}\right|-\frac{g\xi\omega_{1}}{\gamma_{\parallel}}\left[\textrm{Re}\left(\frac{\alpha_{11}}{\beta_{1}}R_{1}\mathcal{L}_{11}\right)\left|E_{1}\right|^{2}+\textrm{Re}\left(\frac{\alpha_{12}}{\beta_{1}}R_{2}\mathcal{L}_{22}\right)\left|E_{2}\right|^{2}\right]\left|E_{1}\right|\\
 & -\textrm{Re}\left[\frac{g\xi\omega_{1}}{\gamma_{\parallel}+2\mathrm{i}\Delta\omega}\frac{\alpha_{12}}{\beta_{1}}\left(\frac{R_{1}}{\beta_{1}}+\frac{R_{2}}{\beta_{2}^{*}}\right)\right]\left|E_{2}\right|^{2}\left|E_{1}\right|,\\
\frac{\mathrm{d}}{\mathrm{d}t}\left|E_{2}\right|\approx\: & \textrm{Re}\left(\frac{g\omega_{2}}{\beta_{2}}R_{2}-\frac{\kappa_{2}}{2}\right)\left|E_{2}\right|-\frac{g\xi\omega_{2}}{\gamma_{\parallel}}\left[\textrm{Re}\left(\frac{\alpha_{22}}{\beta_{2}}R_{2}\mathcal{L}_{22}\right)\left|E_{2}\right|^{2}+\textrm{Re}\left(\frac{\alpha_{12}}{\beta_{2}}R_{1}\mathcal{L}_{11}\right)\left|E_{1}\right|^{2}\right]\left|E_{2}\right|\\
 & -\textrm{Re}\left[\frac{g\xi\omega_{2}}{\gamma_{\parallel}-2\mathrm{i}\Delta\omega}\frac{\alpha_{12}}{\beta_{2}}\left(\frac{R_{1}}{\beta_{1}^{*}}+\frac{R_{2}}{\beta_{2}}\right)\right]\left|E_{1}\right|^{2}\left|E_{2}\right|.\end{aligned}
\label{eq:Ea_final}\end{equation}
\end{widetext}

\subsection{Conditions for bistable lasing: Mode coupling\label{sub:CLA-C}}

With the phase terms dropped, Eqs.~\eqref{eq:Ea_final} represented
in the amplitude form analogous to Eqs.~\eqref{eq:2mode_amplitudes}
can be analyzed following the standard procedure \cite{Siegman}.
The primary parameter that determines the nature of mode competition
is the mode coupling constant \begin{equation}
C=\textrm{Re\,}\theta_{12}\textrm{Re\,}\theta_{21}/\textrm{Re\,}\theta_{11}\textrm{Re\,}\theta_{22},\label{eq:C_definition}\end{equation}
which is the ratio of cross-saturation and self-saturation coefficients.
It is commonly known that the cases of simultaneous two-mode lasing
and bistable lasing are characterized by $C<1$ (weak mode coupling)
and $C>1$ (strong mode coupling), respectively \cite{Siegman,LambLaser}.
Assuming that the pumping does not favor either of the modes so that
$R_{1}=R_{2}\equiv R$, as well as $\omega_{1}\approx\omega_{2}\equiv\omega\gg\Delta\omega$,
we can substitute the explicit form of the coefficients from Eq.~\eqref{eq:Ea_final}
into Eq.~\eqref{eq:C_definition}. As a result, we have found that~$C$
can be factored as \begin{equation}
C=C_{\alpha}C_{\omega}.\label{eq:C_2terms}\end{equation}

The first factor~$C_{\alpha}$, which originates in the spatial hole
burning, has the form\begin{equation}
C_{\alpha}=\frac{\alpha_{12}^{2}}{\alpha_{11}\alpha_{22}}.\label{eq:C_alpha}\end{equation}
In the simplest case when the modes are intensity matched as in Eq.~\eqref{eq:intensities}
so that all $\alpha_{ij}\equiv\alpha$, it follows that~$C_{\alpha}=1$.
Otherwise, it can be proven that $C_{\alpha}\leq1$. The second factor~$C_{\omega}$,
which results from population pulsations and becomes identicaly unity
if those pulsations are neglected, has the form\begin{equation}
C_{\omega}\approx\left(\frac{4\Delta\omega^{2}\left(1-\frac{\gamma_{\parallel}}{\gamma_{\perp}}\right)+2\gamma_{\parallel}^{2}}{\left(4\Delta\omega^{2}+\gamma_{\parallel}^{2}\right)}\right)^{2}+O\left(\frac{\delta_{\omega}^{2}}{\gamma_{\perp}^{2}}\right).\label{eq:C_w}\end{equation}

\begin{figure}
\includegraphics[width=0.5\textwidth]{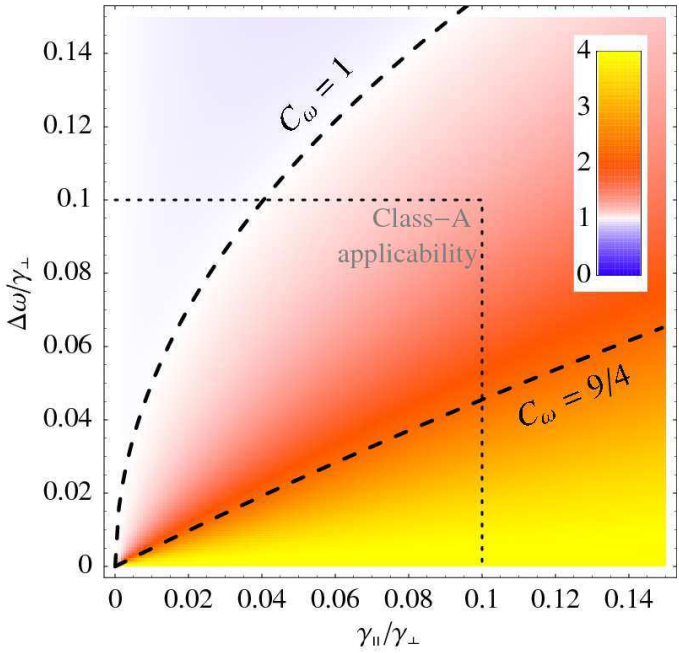}

\caption{(Color online) The dependence of~$C_{\omega}$ in Eq.~\eqref{eq:C_w}
on $\gamma_{\parallel}$~and~$\Delta\omega$. The dashed lines are
the isolines for $C_{\omega}=1$ and $C_{\omega}=9/4$. The dotted
lines approximately mark the applicability limits of class-A equations.
\label{fig:GRAPH_C}}
\end{figure}

The dependence of~$C_{\omega}$ is shown in Fig.~\ref{fig:GRAPH_C}.
We can see that~$C_{\omega}\lesssim4$ for $\Delta\omega\ll\gamma_{\parallel}$
and $C_{\omega}\simeq1$ for $\gamma_{\parallel}<\Delta\omega\ll\gamma_{\perp}$.
The transition between two limiting cases ($C_{\omega}=1$ and $C_{\omega}=4$)
occurs rapidly around $\Delta\omega\simeq\gamma_{\parallel}$. Note
that as $\Delta\omega$~increases, $C_{\omega}$~approaches unity
from below, so there is a critical value $\Delta\omega^{(1)}\approx\sqrt{\gamma_{\perp}\gamma_{\parallel}}/2$
for which $C_{\omega}=1$. Hence, in the ideal case of intensity matched
modes {[}Eq.~\eqref{eq:intensities}] bistability is possible for~$\Delta\omega$
all the way up to $\Delta\omega^{(1)}$. The limiting case of $C=4$
is known to be realized for the ideal case of counterpropagating modes
in ring lasers or modes with orthogonal polarizations, which are fully
intensity matched and have $\Delta\omega\approx0$ \cite{Siegman}.

If, however, the modes are considerably mismatched, then $C_{\omega}$~must
be significantly larger than one to compensate for a small~$C_{\alpha}$
and thus keep the overall mode coupling constant above unity to achieve
bistable lasing. For example, it can be shown that 1D harmonic (e.g.,
longitudinal) modes always have $C_{\alpha}=4/9$ for different frequencies.
This means that the line of critical values for~$\Delta\omega^{(9/4)}\approx\gamma_{\parallel}/2$
up to where bistability is possible lies much deeper than the line
of~$\Delta\omega^{(1)}$ (see Fig.~\ref{fig:GRAPH_C}). Taking into
account that the frequency shift between longitudinal modes is related
to the cavity length as $\Delta\omega_{\textrm{(bulk)}}=\pi c/L$,
one easily obtains the {}``rule of thumb'' for minimum cavity length
of a 1D bistable bulk laser: $L_{\textrm{min}}\simeq2\pi c/\gamma_{\parallel}$.
For realistic laser media, $L_{\textrm{min}}$~is found to be prohibitively
large, from around 2~m for semiconductors and up to 200-300~km for
Nd:YAG \cite{Svelto}. This explains why it is so difficult to achieve
bistable lasing for different-frequency modes in a bulk cavity: unless
the cavity is extraordinary big, $\Delta\omega$~is large enough
to bring~$C_{\omega}$ so close to unity that any intensity mismatch
causes $C_{\alpha}<1$ and brings the laser back into the weak-coupling
(simultaneous lasing) regime. The only notable exception is the case
when the modes are quasi-degenerate with $\Delta\omega\approx0$,
such as counterpropagating modes in ring lasers or modes with orthogonal
polarization, and it is in these special cases that bistability could
indeed be observed.

In a microcavity, however, the modes can be made very nearly intensity
matched by a carefully chosen resonator design (e.g., coupled cavity-based,
see \cite{zhukPRL}). In addition, many designs allow to control the
frequency separation between the modes more or less independently
from other model parameters. This opens up a whole new frequency range
$\Delta\omega^{(9/4)}<\Delta\omega<\Delta\omega^{(1)}$ available
for bistable laser design, which can encompass several orders of magnitude
for~$\Delta\omega$ (see Fig.~\ref{fig:GRAPH_C}). This range becomes
available in microlasers because the possibility to bring the modes
to intensity matching is far greater than in bulk cavities, owing
to a greater variety of cavity shapes and a more complicated nature
of the modes involved.

Finally, from Eqs.~\eqref{eq:Ea_final} one can also see the physical
mechanism of bistable lasing in the class-A case. It is due to the
(oscillatory) component~$W_{12}$ that there is an addition to the
cross-saturation coefficients~$\theta_{ij\neq i}$. Without this
addition, $C$~would simply coincide with~$C_{\alpha}$ and all
possibility for bistable operation would be excluded. Hence, it is
the coherent mode interaction effects such as population pulsations
or four-wave mixing \cite{LambLaser} that make bistability possible.
Incoherent effects (e.g., spatial hole burning, which is only manifest
in~$C_{\alpha}$) can either allow or suppress it. As a result, an
interplay between coherent and incoherent mode interaction processes
is employed to achieve bistable microlaser operation.

\begin{figure*}
\includegraphics[width=0.8\textwidth]{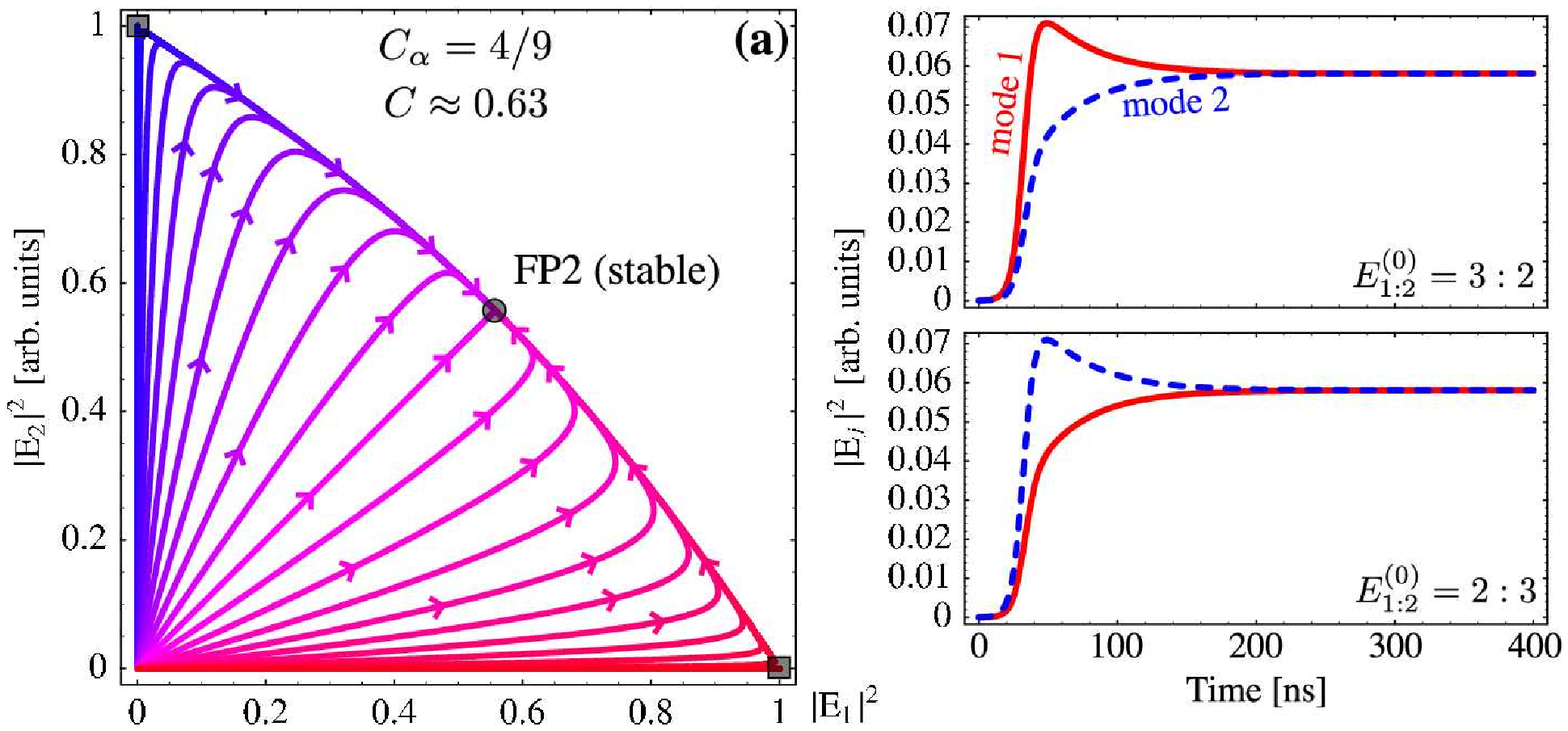}

\bigskip{}

\includegraphics[width=0.8\textwidth]{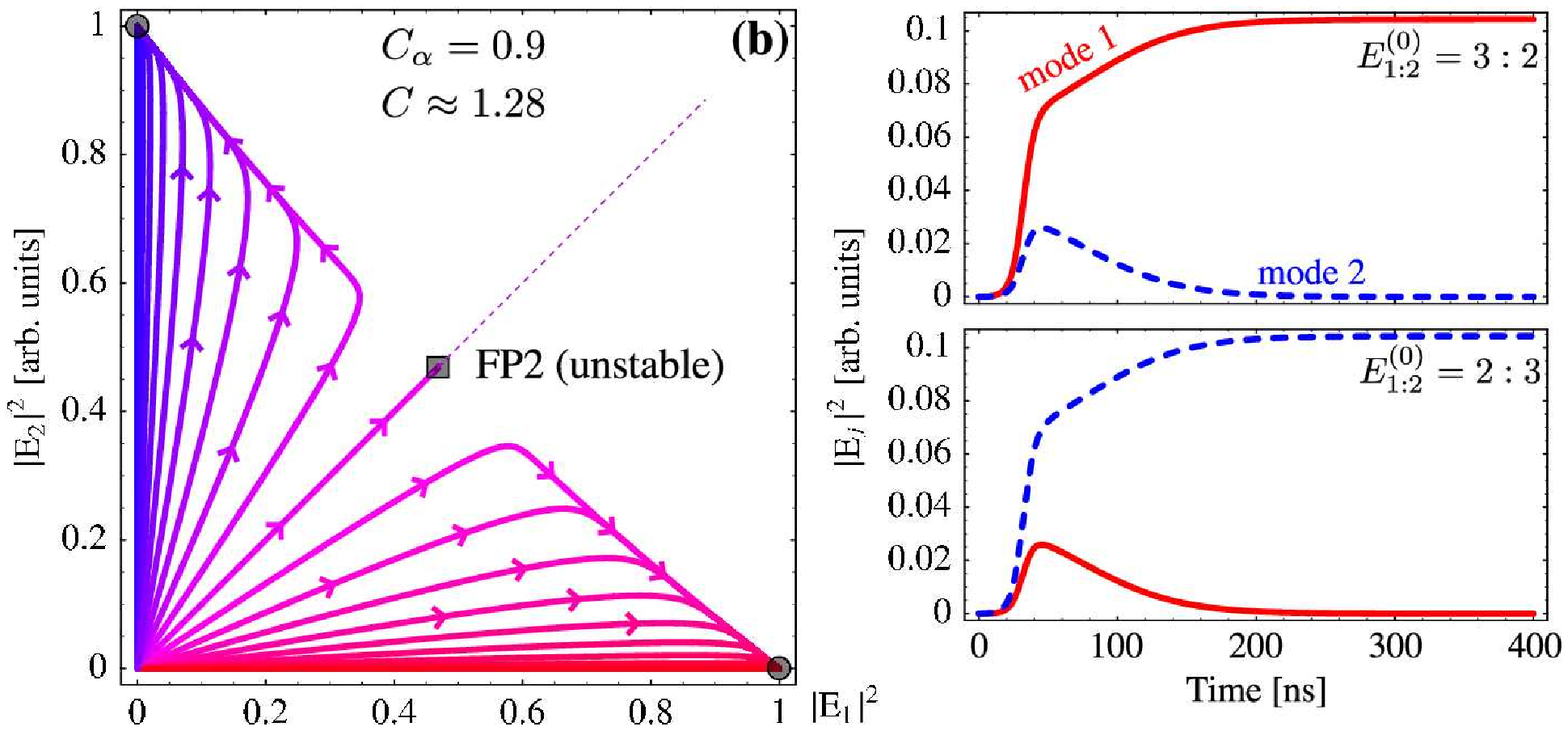}

\caption{(Color online) Flow diagrams of numerical solutions of Eqs.~\eqref{eq:Ea_final}
in the $\left|E_{1}\right|^{2}\textrm{ vs. }\left|E_{2}\right|^{2}$
plane for different initial mode ratios $E_{1:2}^{(0)}=\left|E_{1}(0)\right|:\left|E_{2}(0)\right|$
for \textbf{(a)} poorly matched modes ($C_{\alpha}=4/9$) and \textbf{(b)}
well-matched modes ($C_{\alpha}=0.9$). The other parameters are the
same ($\Delta\omega=\gamma_{\parallel}=10^{-2}\gamma_{\perp}$, $\kappa_{1}=\kappa_{2}=10^{-4}\gamma_{\perp}$),
chosen such that $1<C_{\omega}<9/4$. Solid circles and squares denote
stable and unstable fixed points, respectively. The thin dashed line
denotes the separatrix in the bistable lasing case. The right panels
show the example mode dynamics $\left|E_{1,2}(t)\right|^{2}$ for
two chosen values of~$E_{1:2}^{(0)}$ slightly in favor of each mode.
\label{fig:DIAGRAMS_SYM}}
\end{figure*}

As an example, we have plotted the dynamics of mode amplitudes~$E_{j}(t)$
as a numerical solution of Eqs.~\eqref{eq:Ea_final} for bulk-cavity
($C_{\alpha}=4/9$) vs.~coupled-cavity ($C_{\alpha}=0.9$) modes
(Fig.~\ref{fig:DIAGRAMS_SYM}). Also shown are the temporal flow
diagrams (i.e., projections of the solutions onto the $\left|E_{1}\right|^{2}\textrm{ vs. }\left|E_{2}\right|^{2}$
plane for different initial conditions of the cavity (the ratio $E_{1:2}^{(0)}\equiv\left|E_{1}(0)\right|:\left|E_{2}(0)\right|$).
All other parameters are kept constant, as described in the caption.
If the modes are mismatched (Fig.~\ref{fig:DIAGRAMS_SYM}a) and $C<1$,
the laser saturates to the two-mode simultaneous lasing ($\left|E_{1}\right|=\left|E_{2}\right|=\textrm{const}$)
regardless of the initial conditions. Only this fixed point is stable.
However, if the modes are well matched (Fig.~\ref{fig:DIAGRAMS_SYM}b)
so that $C>1$, the laser saturates to a single-mode lasing as the
initially stronger mode quenches its weaker counterpart and becomes
dominant. There are two stable fixed points on the diagram: $\left|E_{1}\right|=\textrm{const},\;\left|E_{2}\right|=0$
and $\left|E_{1}\right|=0,\;\left|E_{2}\right|=\textrm{const}$. The
previously stable fixed point becomes unstable, and the line $\left|E_{1}\right|=\left|E_{2}\right|$
marks the separatrix between the stable points' domains of attraction.
The mode that has an advantage in the beginning determines the domain
of attraction for the system and hence the fixed point the system
will converge to, as the separatrix cannot be transcended without
an external influence. These examples show that bistable lasing is
possible in microlasers in such cases where only two-mode simultaneous
lasing can be observed for bulk-cavity modes.

\subsection{Conditions for bistable lasing: Mode mismatch\label{sub:CLA-I12}}

Up to now, we assumed that none of the modes is favored either by
the cavity or by the gain, i.e., $\kappa_{1}=\kappa_{2}$, $R_{1}=R_{2}$,
$\alpha_{11}=\alpha_{12}$, and $\delta_{\omega}=0$. In this case,
as seen in Fig.~\ref{fig:DIAGRAMS_SYM}, the two-mode lasing fixed
point (labeled FP2), whether stable or unstable, is characterized
by $\left|E_{1}\right|=\left|E_{2}\right|$. This means, on the one
hand, that in the simultaneous-lasing case both modes lase with equal
intensity (Fig.~\ref{fig:DIAGRAMS_SYM}a), and on the other hand,
that in the bistable regime even a slight edge given to either mode
in terms of initial conditions will bring this mode to lase. It is
equally easy to {}``select'' or {}``switch'' either mode by locking
into it \cite{zhukPRL,zhukPSS}. This is illustrated in Fig.~\ref{fig:DIAGRAMS_SYM}b
by the fact that each stable fixed point has an equally large domain
of attraction. 

\begin{figure*}
\includegraphics[width=0.8\textwidth]{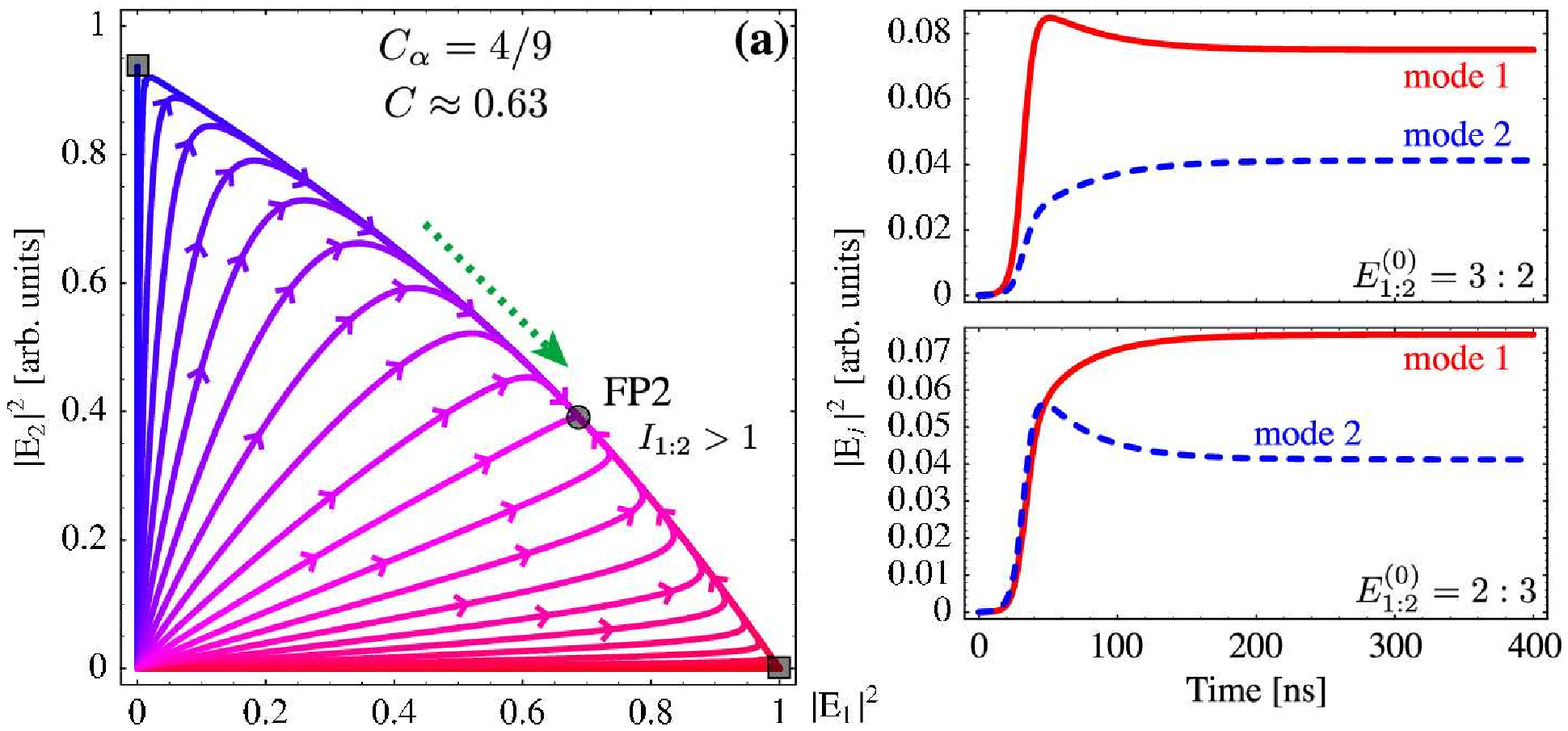}

\bigskip{}

\includegraphics[width=0.8\textwidth]{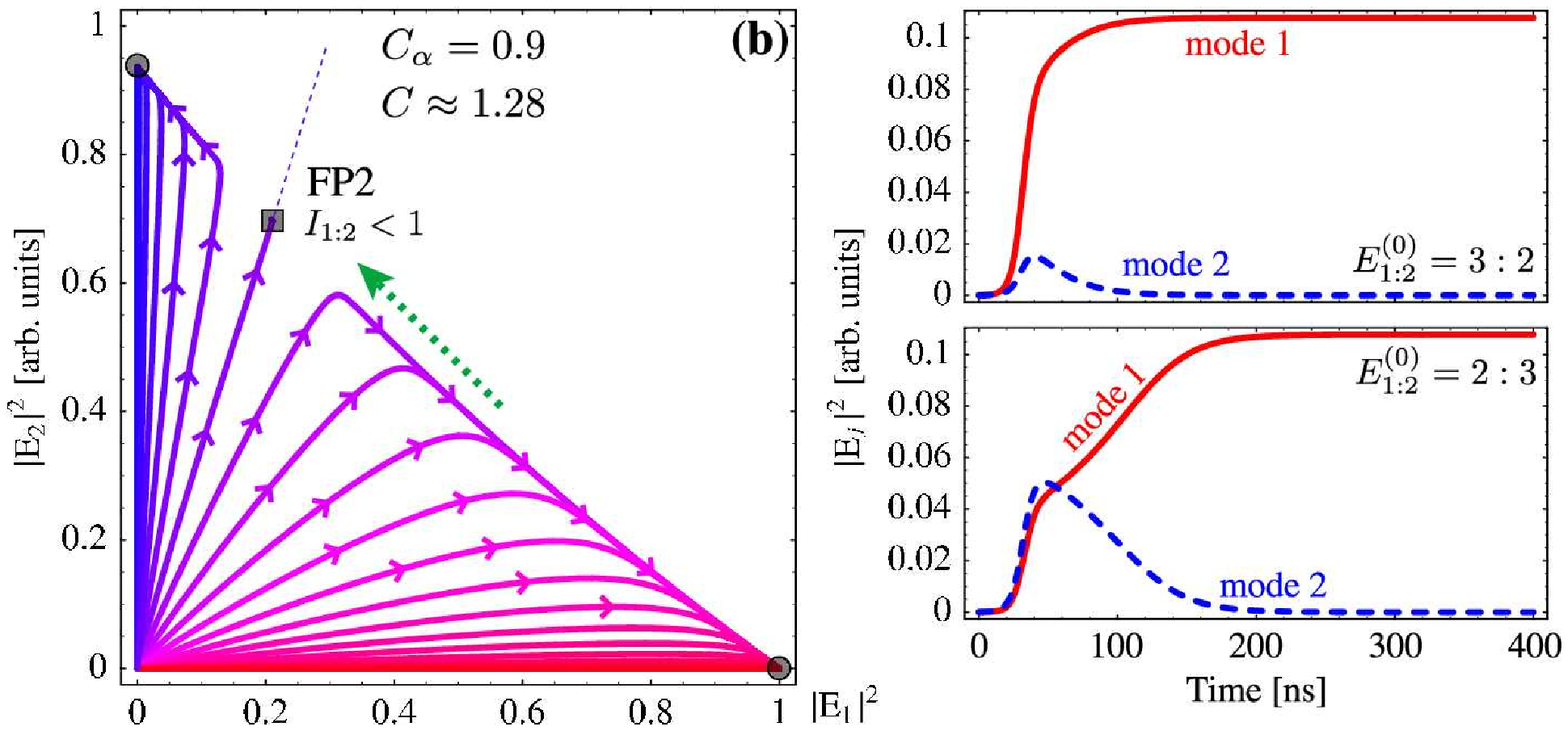}

\caption{(Color online) Same as Fig.~\ref{fig:DIAGRAMS_SYM} for mismatched
mode decay rates ($\kappa_{1}<\kappa_{2}$, $\delta_{\kappa}=-0.42\%$)
in favor of the first mode. The dotted arrows show the shift of the
two-mode fixed point FP2. \label{fig:DIAGRAMS_ASYMM}}
\end{figure*}

In a more general case, the ratio between mode intensities at FP2
$I_{1:2}\equiv\left|E_{1}\right|^{2}/\left|E_{2}\right|^{2}$ will
change to reflect an advantage given to either of the modes. For example,
even a slight mismatch in the mode $Q$-factors causes $I_{1:2}$
to deviate from unity (Fig.~\ref{fig:DIAGRAMS_ASYMM}). Similar to
the explanation given above, this may mean two things. In the simultaneous-lasing
case, it simply means that once the laser achieves saturation, one
mode has a greater amplitude than the other, e.g., $\left|E_{1}\right|>\left|E_{2}\right|$
for $I_{1:2}>1$ (Fig.~\ref{fig:DIAGRAMS_ASYMM}a). In the bistable
case, it means that the domains of attraction for the two modes change
their size in phase space (Fig.~\ref{fig:DIAGRAMS_ASYMM}b). If for
example $I_{1:2}<1$, then the domain of attraction for Mode~1 becomes
larger, so Mode~1 is {}``in favor'' as a result. In the bistable
regime, a shifted FP2 means that the mode with a smaller domain of
attraction is out of favor and thus harder to bring to lasing. For
example, if FP2 is placed symmetrically, initial mode amplitude ratios~$E_{1:2}^{(0)}$
of 3:2 and 2:3 bring the first and the second mode to lasing, respectively
(Fig.~\ref{fig:DIAGRAMS_SYM}b). For asymmetrically placed FP2, the
same two cases for initial condition both result in the lasing of
the first mode (Fig.~\ref{fig:DIAGRAMS_ASYMM}b). To be able to target
the smaller domain, one has to excite the out-of-favor mode exclusively,
which might be difficult experimentally. Hence we will further aim
at finding the manifold of the system parameters for which $I_{1:2}=1$.

Whenever $C\neq1$, the general expression for the mode intensity
ratio at FP2~$I_{1:2}$ can be written as \cite{Siegman}\begin{equation}
I_{1:2}=\frac{\textrm{Re\,}\rho_{1}\textrm{Re\,}\theta_{22}-\textrm{Re\,}\rho_{2}\textrm{Re\,}\theta_{12}}{\textrm{Re\,}\rho_{2}\textrm{Re\,}\theta_{11}-\textrm{Re\,}\rho_{1}\textrm{Re\,}\theta_{21}}.\label{eq:I12_definition}\end{equation}
By substituting the coefficients in Eqs.~\eqref{eq:Ea_final} one
can obtain an explicit analytic expression for~$I_{1:2}$. Unfortunately,
this general expression is very bulky and we will first investigate
its behavior in several simplified cases. Let us introduce the perturbations
in the form\begin{equation}
\kappa_{1,2}\equiv\kappa(1\pm\delta_{\kappa}),\quad R_{1,2}\equiv R(1\pm\delta_{\alpha}),\label{eq:a12_k12_perturbation}\end{equation}
from where it follows {[}see Eqs.~\eqref{eq:W_comps} and~\eqref{eq:alpha_comps}]
that $\alpha_{11,22}=\alpha(1\pm\delta_{\alpha})^{2}$. Now if $\delta_{\omega}=\delta_{\alpha}=0$,
$\delta_{\kappa}\neq0$, $I_{1:2}$ is given by:\begin{equation}
I_{1:2}^{(\kappa)}=\frac{a_{\kappa}+b_{\kappa}\delta_{\kappa}}{a_{\kappa}+b_{\kappa}\delta_{\kappa}}.\label{eq:I12_Dkappa}\end{equation}
Likewise if $\delta_{\omega}=\delta_{\kappa}=0$, $\delta_{\alpha}\neq0$,
then the expression is somewhat more complicated and reads:\begin{equation}
I_{1:2}^{(\alpha)}=\frac{\left(a_{\alpha}+c_{\alpha}\delta_{\alpha}^{2}+e_{\alpha}\delta_{\alpha}^{4}\right)+\left(b_{\alpha}+d_{\alpha}\delta_{\alpha}^{2}\right)\delta_{\alpha}}{\left(a_{\alpha}+c_{\alpha}\delta_{\alpha}^{2}+e_{\alpha}\delta_{\alpha}^{4}\right)-\left(b_{\alpha}+d_{\alpha}\delta_{\alpha}^{2}\right)\delta_{\alpha}}.\label{eq:I12_Dalpha}\end{equation}
Finally, if if $\delta_{\alpha}=\delta_{\kappa}=0$, $\delta_{\omega}\neq0$,
then \begin{equation}
I_{1:2}^{(\omega)}=\frac{\left(a_{\omega}+c_{\omega}\delta_{\omega}^{2}+e_{\omega}\delta_{\omega}^{4}\right)+\left(b_{\omega}+d_{\omega}\delta_{\omega}^{2}+f_{\omega}\delta_{\omega}^{4}\right)\delta_{\omega}}{\left(a_{\omega}+c_{\omega}\delta_{\omega}^{2}+e_{\omega}\delta_{\omega}^{4}\right)-\left(b_{\omega}+d_{\omega}\delta_{\omega}^{2}+f_{\omega}\delta_{\omega}^{4}\right)\delta_{\omega}}.\label{eq:I12_delta}\end{equation}
The coefficients in Eqs.~\eqref{eq:I12_Dkappa}--\eqref{eq:I12_delta}
are complicated polynomial functions of the dynamical parameters $\gamma_{\perp}$~and~$\gamma_{\parallel}$,
the intermode frequency separation~$\Delta\omega$, the measure of
mode intensity mismatch $\nu\equiv\alpha_{12}/\alpha$ (which ranges
from 0 to a maximum value of $1-\delta_{\alpha}^{2}$ so that $C_{\alpha}\leq1$),
and the pumping rate normalized to the threshold pumping $R_{\textrm{thr}}\equiv2Rg\omega/(\gamma_{\perp}\kappa)$
\footnote{Note that from the way the class-A equations were constructed, $R/R_{\textrm{thr}}$
cannot exceed one significantly. Numerical analysis shows that the
mode dynamics no longer change if $R/R_{\textrm{thr}}$ is increased
beyond 10, which is roughly where the near-threshold iterative expansion
that yields the solution in the form of Eq.~\eqref{eq:Wa_ansatz}
ceases to be applicable.%
}. Note that $\kappa$~itself does not enter these equations explicitly.
It does, however, impose a limitation $\Delta\omega>10^{-2}\kappa$
so that the phase terms in Eqs.~\eqref{eq:Ea_relaxed} can be averaged
out. 

From the structure of Eqs.~\eqref{eq:I12_Dkappa}--\eqref{eq:I12_delta}
one can see that $I_{1:2}=1$ for $\delta_{\alpha}=\delta_{\kappa}=\delta_{\omega}=0$,
as should be expected. If any one of the perturbation parameters ($\delta_{\omega,\kappa,\alpha}$
collectively referred to as~$\delta$) is non-zero, $I_{1:2}$~deviates
from unity. Obviously, changing the sign of all non-zero~$\delta$
causes $I_{1:2}\to1/I_{1:2}$. If favoring one of the modes (by any
means) results in a certain asymmetry in lasing quantified through
a non-unity $I_{1:2}$, then favoring the other mode in the same way
and by the same amount naturally causes the same asymmetry with respect
to the other mode %
\footnote{There is no simple way to tell if the coefficients in Eqs.~\eqref{eq:I12_Dkappa}--\eqref{eq:I12_delta}
are positive or negative for a given set of parameters. For instance,
whenever the coefficients $a_{\kappa}$~or~$b_{\kappa}$ change
sign in Eq.~\eqref{eq:I12_Dkappa}, similar~$\delta_{\kappa}$ will
cause an opposite shift in~$I_{1:2}$.%
}. This suggests that one can choose \emph{more than one}~$\delta$
to be non-zero in such a way that the shifts of FP2 caused by individual
perturbations would cancel each other out. As a result, one could
achieve the resulting~$I_{1:2}$ equal to or close to unity, and
the restrictions on the initial conditions would be lifted.

\begin{figure}
\includegraphics[width=0.3\textwidth]{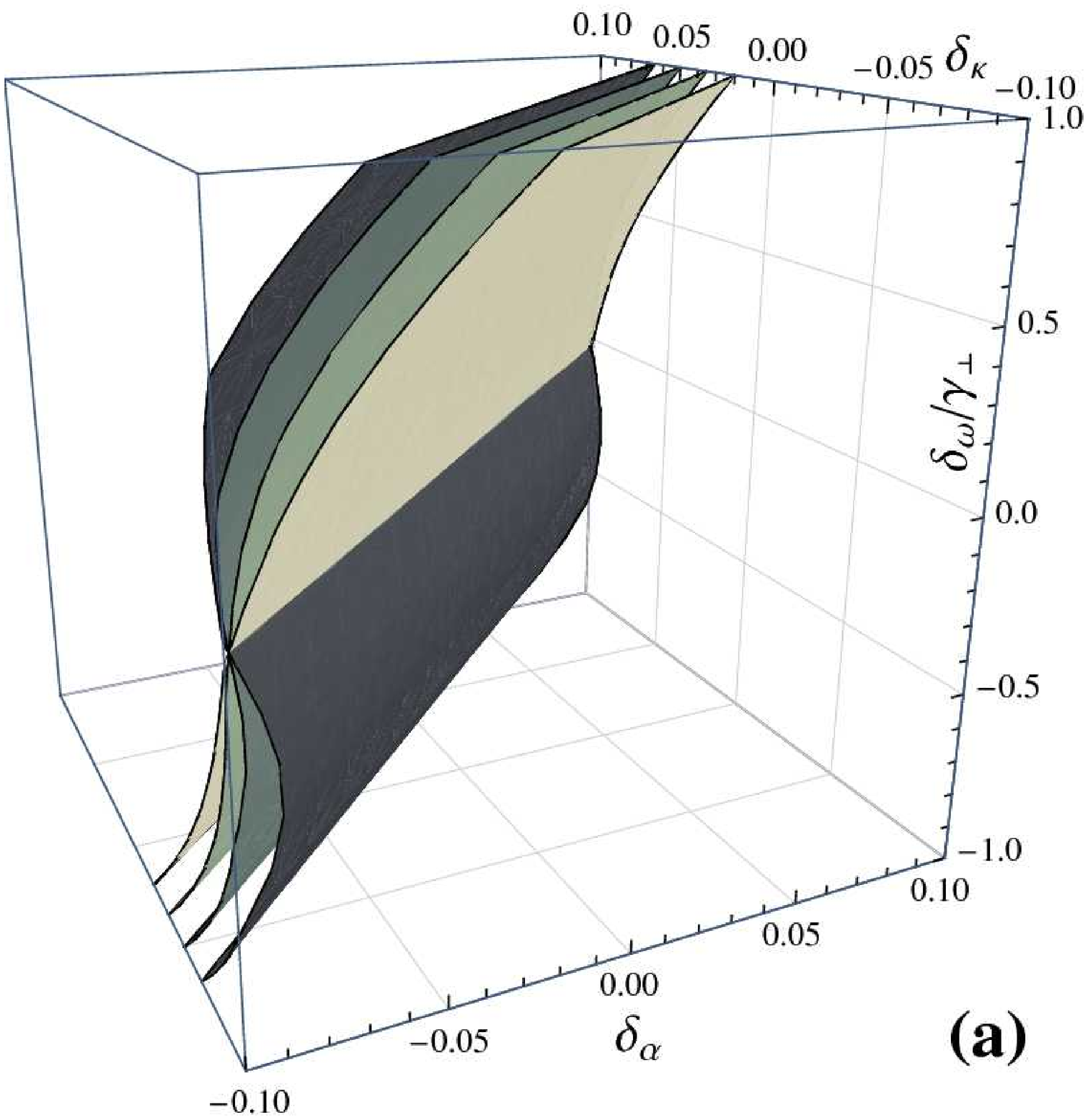}\hfill{}\includegraphics[width=0.3\textwidth]{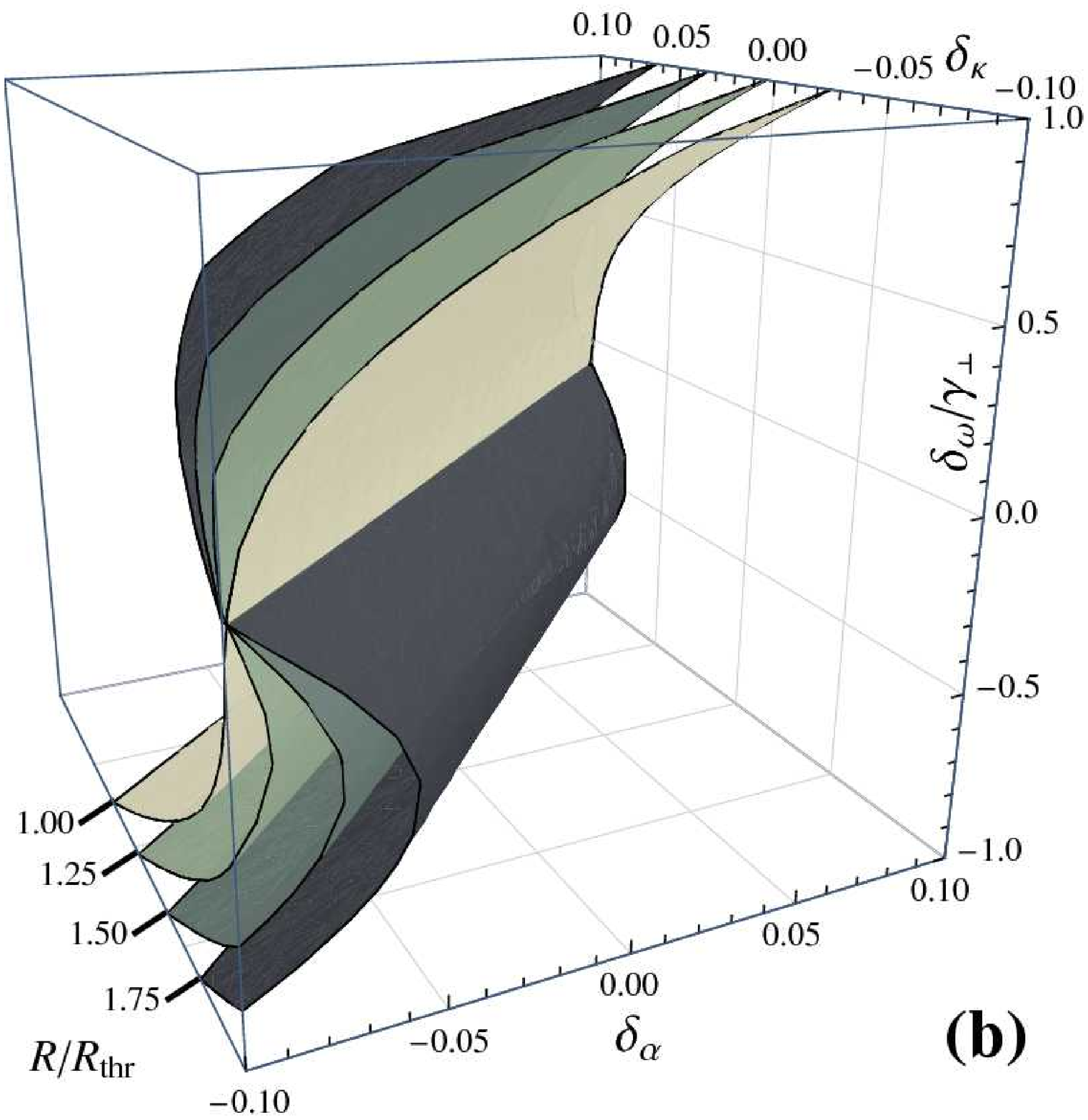}\hfill{}\includegraphics[width=0.3\textwidth]{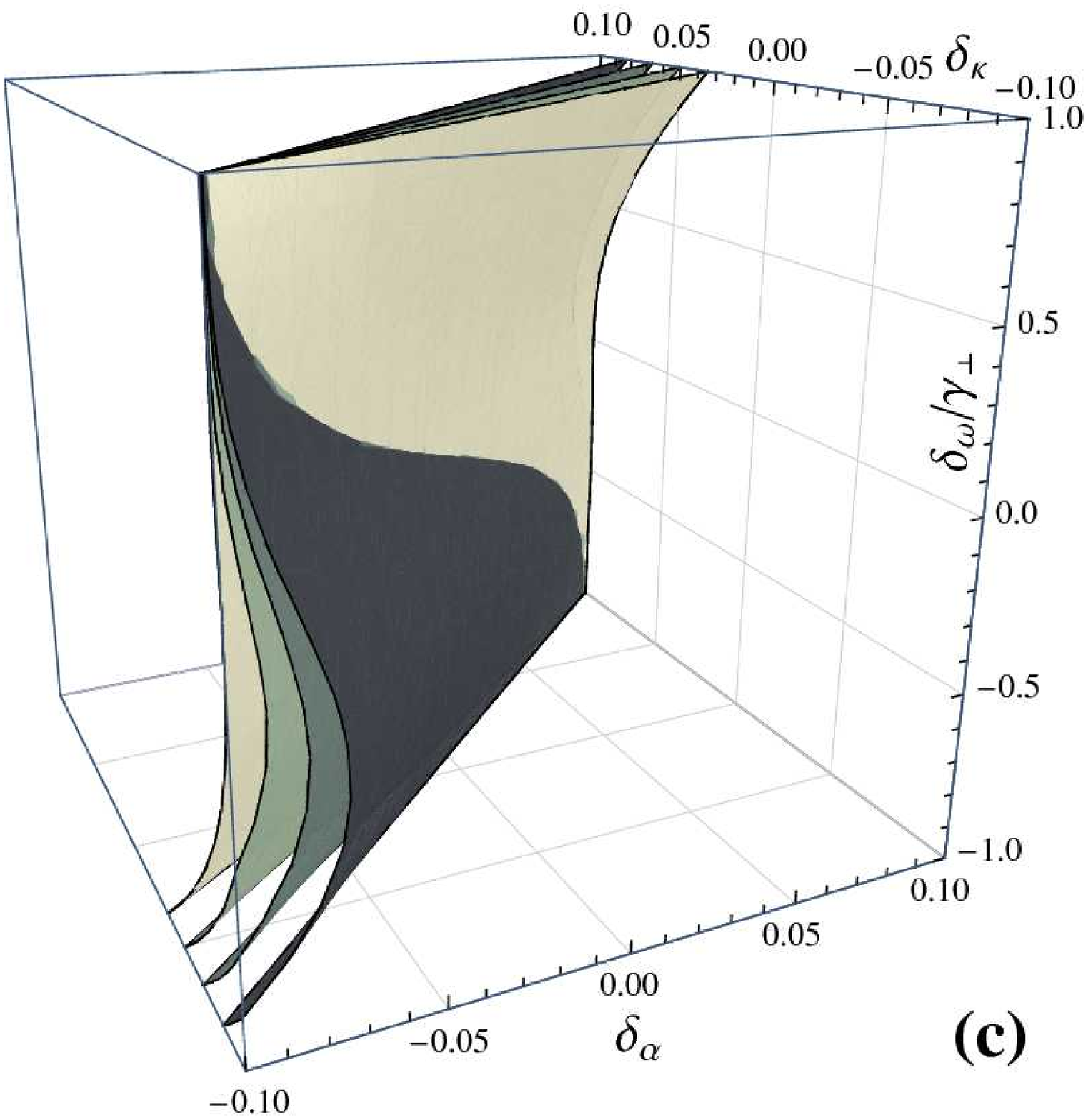}

\caption{(Color online) Manifolds of points $I_{1:2}=1$ in the 3D perturbation
space $(\delta_{\omega};\delta_{\kappa};\delta_{\alpha})$ for \textbf{(a)}
$C=1.10\gtrsim1$, \textbf{(b)} $C=2.06\simeq2$, and \textbf{(c)}
$C=3.84\lesssim4$. The four surfaces correspond to four values of
the pumping rate ($R/R_{\textrm{thr}}=1,1.25,1.5,1.75$), as indicated
in the panel (b). \label{fig:I12_ZERO}}
\end{figure}

Fig.~\ref{fig:I12_ZERO} shows the manifold of the points $I_{1:2}=1$
in the 3D perturbation space $(\delta_{\omega};\delta_{\kappa};\delta_{\alpha})$
for different parameters as a numerical solution of Eq.~\eqref{eq:I12_definition}.
We can see that this manifold is an open surface. Hence, if a mismatch
in one respect is unavoidable, it can be compensated for by engineering
the other two perturbation parameters. Note that in the $(\delta_{\kappa};\delta_{\alpha})$
plane the mismatch compensation ($I_{1:2}=1$) is achieved when $\delta_{\kappa}\approx\delta_{\alpha}$.
This is easily understood if one remembers that the linear terms in
Eqs.~\eqref{eq:Ea_relaxed} have the structure $\rho_{j}\sim\zeta R_{j}-\kappa_{j}=\zeta R(1\pm\delta_{\alpha})-\kappa(1\pm\delta_{\kappa})$.
On the other hand, in the $(\delta_{\omega};\delta_{\kappa})$ plane,
compensation is generally achieved for the oppositely-signed $\delta_{\omega}$~and~$\delta_{\kappa}$.
This is in agreement with an intuitive guess that, e.g., $\delta_{\kappa}<0$
($\kappa_{1}<\kappa_{2}$) and $\delta_{\omega}<0$ (the gain frequency
$\omega_{a}<\omega_{0}$ is closer to~$\omega_{1}$ than to~$\omega_{2}$,
see Fig.~\ref{fig:FREQUENCIES}) both give an edge to the first mode,
so oppositely-signed $\delta$ are needed to maintain balance. However,
in the vicinity of the origin the surface can be folded, so that it
crosses the origin with the opposite slope and compensation is achieved
when $\delta_{\omega}$ and $\delta_{\kappa}$ have the same sign.
Since perturbations $\delta_{\omega}$,~$\delta_{\alpha}$, and~$\delta_{\kappa}$
can have different physical origin and can be varied more or less
independently by a proper choice of a gain medium and a cavity configuration,
one can deliberately engineer a microlaser to achieve bistable operation
even if the idealized, unperturbed case is difficult to realize experimentally.
An example of such compensation is changing the mode frequencies with
respect to gain (which can be done straightforwardly just by scaling
the cavity) to help offset the difference in mode $Q$-factors, as
shown numerically in our earlier work \cite{zhukPSS}.

To achieve a fully symmetric placement of FP2, one needs to bring
three perturbation parameters into a relation. Because all these parameters
show only an indirect dependency on the cavity design and/or gain
medium choice, the precise control of them may still be a challenging
task. Hence, it is worthwhile to investigate to what extent the relations
for ideal compensation can be violated so that bistable operation
is still possible (albeit, as shown above, at the cost of stricter
requirements on the initial conditions). In terms of Fig.~\ref{fig:I12_ZERO},
that means how far one can deviate from the $I_{1:2}=1$ surface and
still lase into either of the modes on demand.

From Eqs.~\eqref{eq:I12_Dkappa}--\eqref{eq:I12_delta} one sees
that a sufficiently high value of any~$\delta$ will cause either
the numerator or the denominator in~$I_{1:2}$ to approach zero.
On the flow diagram, this corresponds to the FP2 meeting the coordinate
axes. Increasing~$\delta$ further causes~$I_{1:2}$ to become negative.
The FP2 vanishes and the system finds itself in the single mode lasing
regime (see \cite{Siegman}). That sets an upper limit for any $\left|\delta\right|$
beyond which no bistable lasing is possible any more.

\begin{figure}
\includegraphics[width=0.3\textwidth]{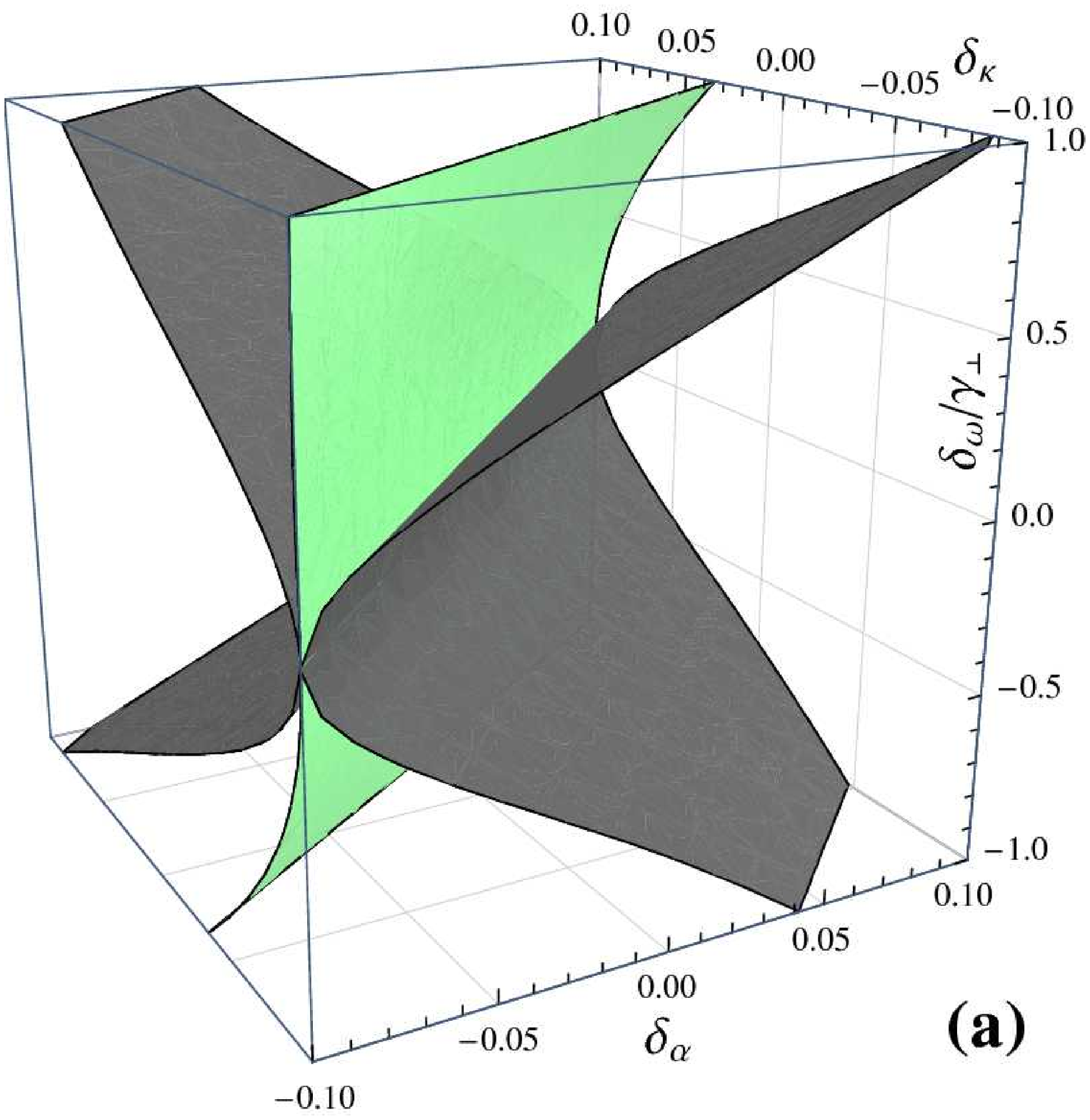}\hfill{}\includegraphics[width=0.3\textwidth]{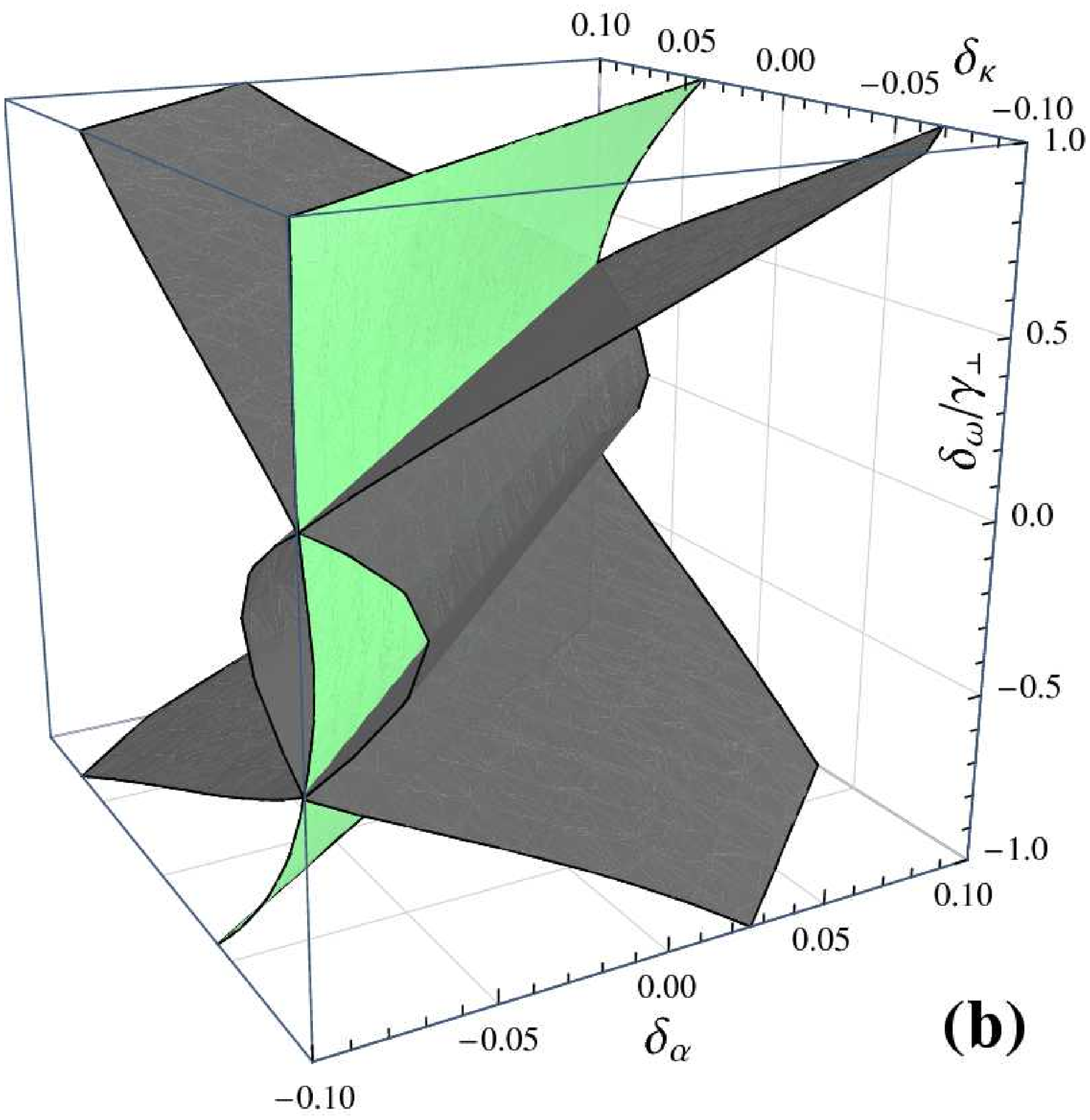}\hfill{}\includegraphics[width=0.3\textwidth]{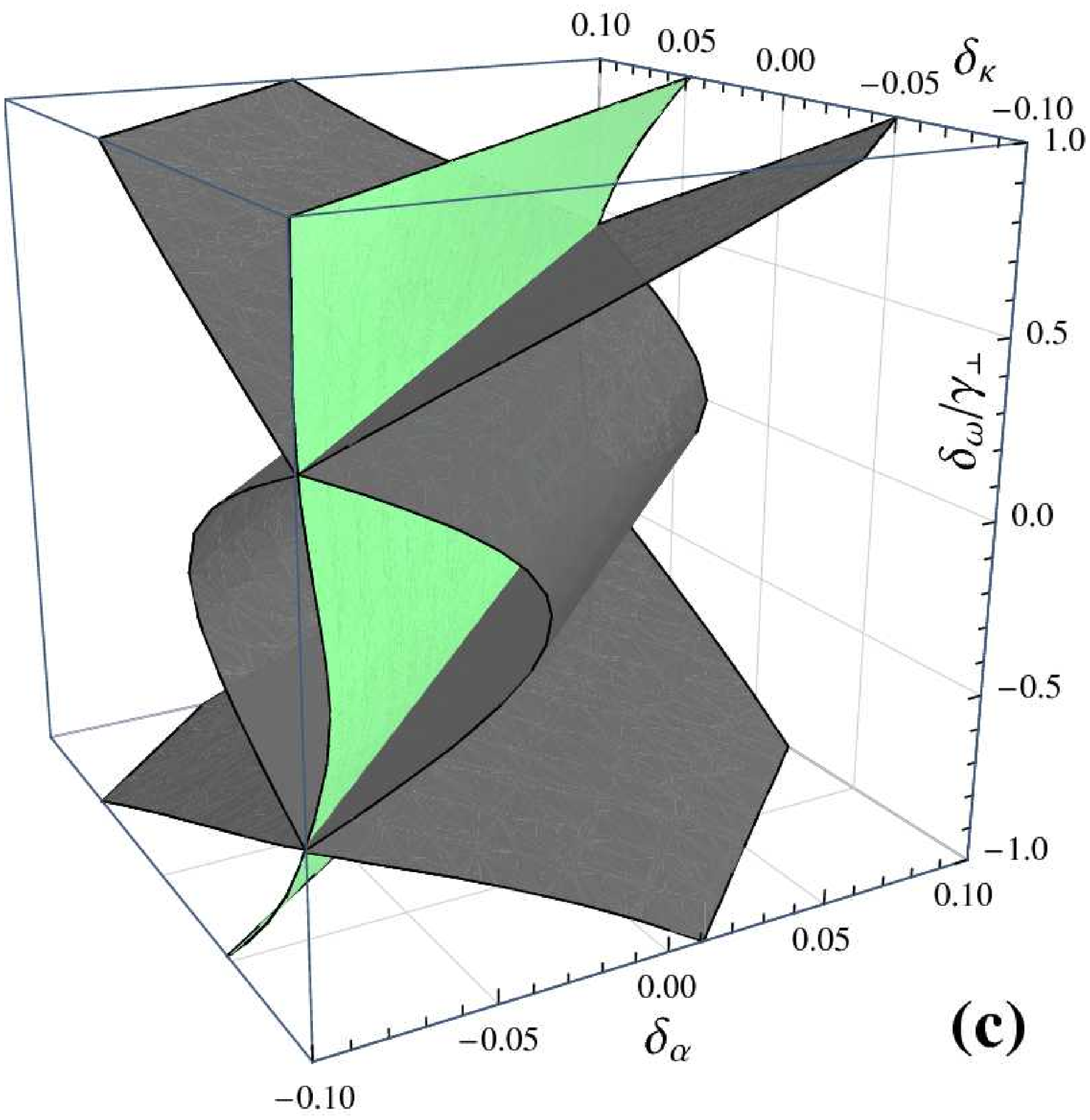}

\caption{(Color online) Boundaries of the FP2 existence domain $I_{1:2}>0$
(dark gray) and the $I_{1:2}=1$ surface lying inside that domain
(light green) for $C\lesssim4$ (as in Fig.~\ref{fig:I12_ZERO}c):
(a) at threshold ($R/R_{\textrm{thr}}=1$), (b) 10\% above threshold
($R/R_{\textrm{thr}}=1.2$), and (c) 20\% above threshold ($R/R_{\textrm{thr}}=1.2$).\label{fig:I12_DOMAIN_R}}
\end{figure}

\begin{figure}
\includegraphics[width=0.3\textwidth]{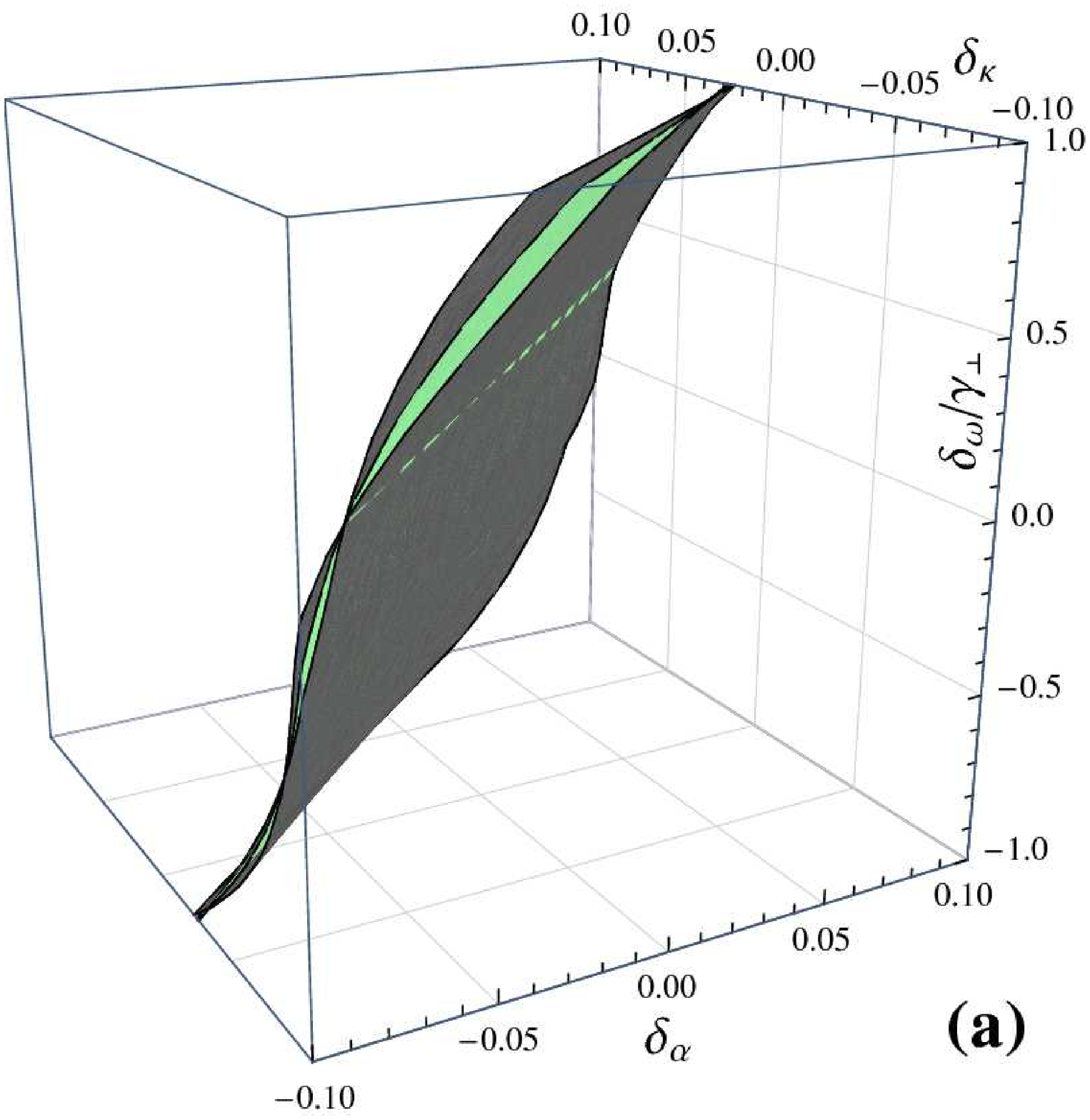}\hfill{}\includegraphics[width=0.3\textwidth]{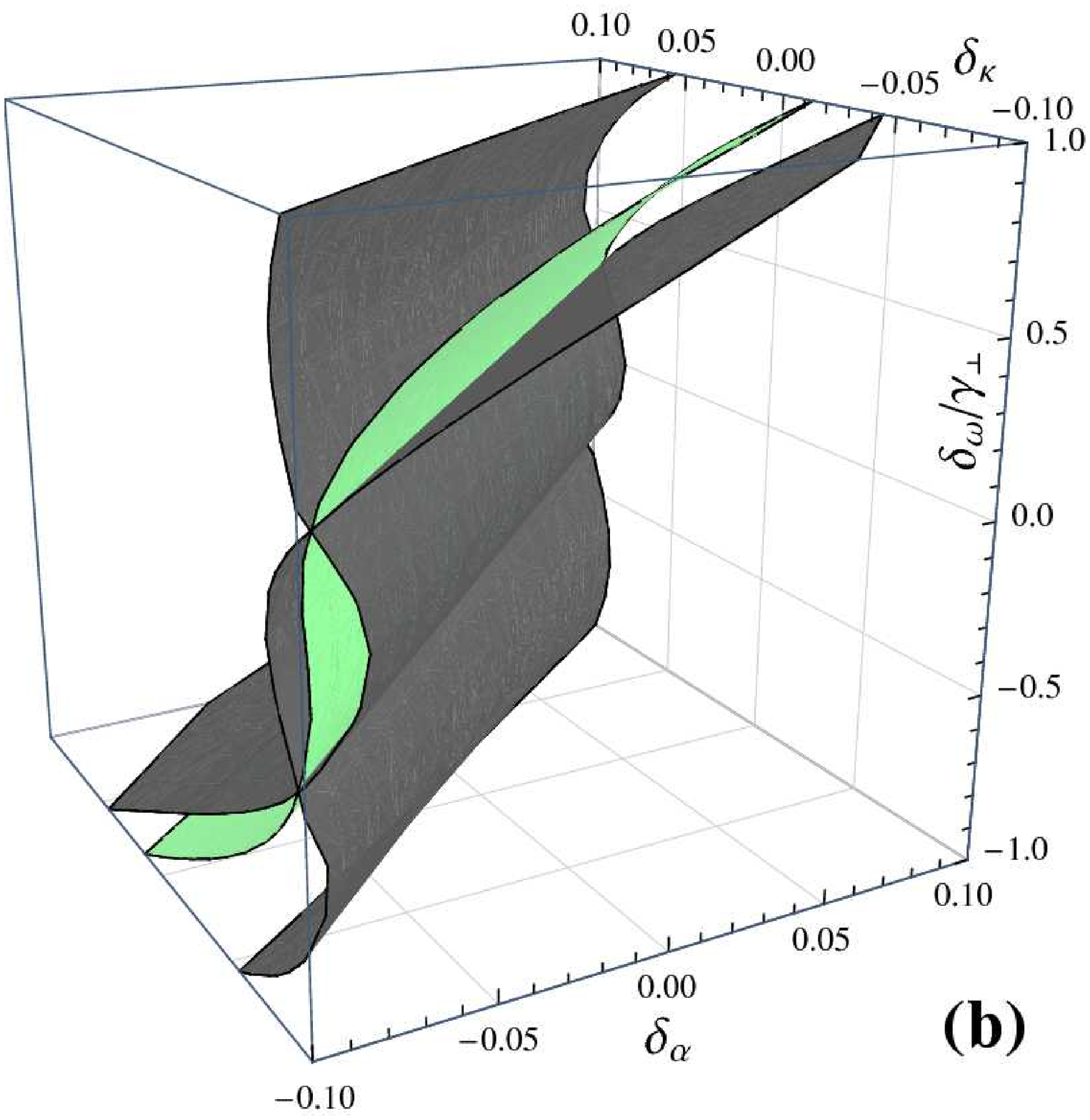}\hfill{}\includegraphics[width=0.3\textwidth]{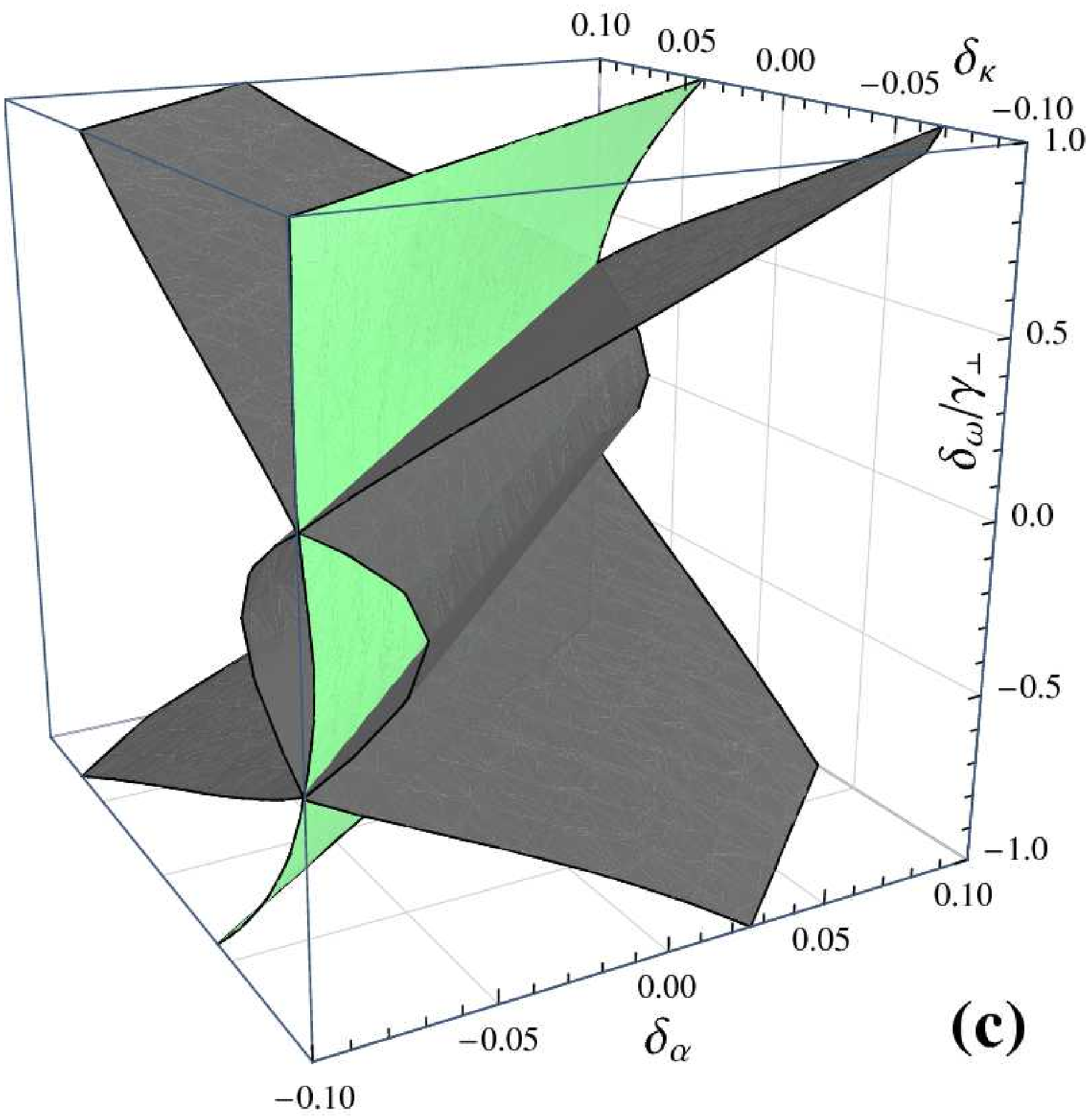}

\caption{(Color online) Same as Fig.~\ref{fig:I12_DOMAIN_R} for constant~$R$
at 10\% above threshold and (a) $C\gtrsim1$, (b) $C\simeq2$, (c)
$C\lesssim4$, corresponding to the cases in Fig.~\ref{fig:I12_ZERO}a--c.\label{fig:I12_DOMAIN_C}}
\end{figure}

More generally, the domain in space $(\delta_{\omega};\delta_{\kappa};\delta_{\alpha})$
where~$I_{1:2}>0$ comprises the possible perturbation parameter
window where both modes can lase (either simultaneously or subject
to bistability-induced switching, as depends on~$C$). This domain,
called the FP2 existence domain, is shown in Figs.~\ref{fig:I12_DOMAIN_R}--\ref{fig:I12_DOMAIN_C}.
The existence domain, bounded by the surfaces defined by $I_{1:2}=0$
and $I_{1:2}=\infty$ is seen to surround the {}``perfect matching''
surface $I_{1:2}=1$. The domain boundaries appear to slide inwards
as the pumping rate increases (Fig.~\ref{fig:I12_DOMAIN_R}), which
enlarges the FP2 existence domain around the point $\delta=0$. Also,
the domain shrinks rapidly as the boundaries close around $I_{1:2}=1$
when $C$~approaches unity (Fig.~\ref{fig:I12_DOMAIN_C}). The latter
can be intuitively understood because $C\approx1$ represents a delicately
balanced system, so that even a slight mismatch is enough to throw
the system heavily out of balance. Such a property is clearly a misfortune
for the microcavity-specific bistability range reported above, as
it relies on the situation when~$C_{\omega}$ exceeds unity only
slightly. However, increasing the pumping appears to counteract this
disadvantage, at least for smaller~$\delta_{\omega}$ (see Fig.~\ref{fig:I12_DOMAIN_R}).
We believe that it is this effect that enabled us to observe bistability
in earlier numerical simulations \cite{zhukPRL,zhukPSS} involving
the laser operating highly above threshold.

The practical conclusion to this section is that there are two theoretical
requirements needed to achieve bistable lasing. In the first place,
FP2 needs to exist on the flow diagram, as imposed by $I_{1:2}>0$.
In the second place, once FP2 exists, the mode coupling constant must
exceed unity ($C>1$), as discussed before. First (Sec.~\ref{sub:CLA-C}),
we have shown that in comparison to bulk-cavity lasers microlasers
exhibit a much wider parameter window characterized by~$C>1$, because
the microcavity modes can better fulfill the intensity matching condition~\eqref{eq:intensities}.
Secondly (Sec.~\ref{sub:CLA-I12}), we have shown that there is an
extended domain in the 3D perturbation space $(\delta_{\omega};\delta_{\kappa};\delta_{\alpha})$
where $I_{1:2}>0$. Inside this domain, the closer $I_{1:2}$~is
to unity, the easier it is to realize bistability-based laser mode
switching experimentally. We have shown that $I_{1:2}$ can be brought
close to~1 by choosing a combination of perturbation parameters that
would compensate each other's advantage given to either mode.

\section{Class-B/C microlasers \label{sec:CLASS-BC}}

The elegance of the class-A case considered in the previous section
is that Eqs.~\eqref{eq:Ea_relaxed} can be subject to analytical
investigation based on a comparison with Eqs.~\eqref{eq:2mode_competition}
\cite{Siegman}. Once a laser with a more complicated dynamics needs
to be examined, more complicated systems of equations {[}six equations~\eqref{eq:Eb}--\eqref{eq:Wb}
for class-B or eight equations~\eqref{eq:Ec}--\eqref{eq:Wc} for
class-C] need to be dealt with. Although attempts at analytical investigation
of class-B equations are known (e.g., a near-threshold expansion of
population inversion as proposed in \cite{mcZehnle}), only numerical
solution seems to be applicable in the general case when no specific
assumptions on the cavity or mode geometry are implied. Since all
the equations are ordinary differential, such a numerical solution
can be carried out with relative ease -- the computational demands
are far lower than a direct numerical integration of the Maxwell-Bloch
equations by means of an FDTD-like scheme \cite{zhukFDTD}.

A systematic investigation of class-B/C microlasers would be too lengthy
to include in the present paper and will be the subject of a forthcoming
publication. In this section we will outline the main differences
in the behavior of such lasers compared to the previously studied
class-A case as regards bistable lasing.

We begin with a comparison of the laser classes in the near-threshold
regime. As should be expected, the solutions for all classes display
full coincidence if the class-A approximation $\gamma_{\perp}\gg\gamma_{\parallel}\gg\kappa$
holds (note that this condition is rather restrictive in microlasers,
requiring a careful choice of the gain medium as well as the cavity
design). The mode dynamics $E_{j}(t)$ start to exhibit differences
whenever $\gamma_{\parallel}$~or~$\kappa$ are increased out of
the class-A approximation. The differences, however, are relatively
minor, manifesting themselves mainly in the character of the transition
process. In most cases, the mode coupling constant~$C$ as defined
for the class-A in Eqs.~\eqref{eq:C_2terms}--\eqref{eq:C_w} continues
to predict the laser dynamics correctly ($C<1$: simultaneous lasing,
$C>1$: bistability) even outside its strict range of applicability,
although the behavior of~$E_{j}(t)$ can be quite different during
the transition period.

As discussed above, the class-B equations \eqref{eq:Eb}--\eqref{eq:Wb}
do not involve a near-threshold approximation, it becomes possible
to consider a greater range of pumping rates, including regimes far
above threshold, which are often left out of the picture in a construction
of a multimode laser model \cite{mcHodges}. Comparison of the numerical
results for class-B vs.~class-C equations show that as long as the
class-B prerequisites $\gamma_{\perp}\gg\gamma_{\parallel},\kappa_{j}$
hold, the results are similar, unless the condition $\gamma_{\perp}\gg\Delta\omega$
is violated. This agrees well with the earlier discussions in Section~\ref{sub:EQS-B}.
The differences appear not to be qualitative, but quantitative only,
manifesting in the exact shape of the $\left|E_{j}(t)\right|$ dependence.
The overall outcome of the mode interaction largely remains the same.
 To summarize, the main effect of the class-A to class-B transition
in the context of studying bistable lasing is the inclusion of larger
pumping rates~$R$, while the main effect of the class-B to class-C
transition is the inclusion of larger frequency mode separations~$\Delta\omega$.

The increase of the pumping rate in a class-B laser is known to change
the saturation character of the mode amplitudes. The non-instantaneous
relaxation of the population inversion with respect to the cavity
field gives rise to spiking (for smaller~$R$) or relaxation oscillations
(for greater~$R$) in the dependence $E_{j}(t)$. A still stronger
pumping (several orders of magnitude above threshold) causes the oscillations
to vanish, as reported in an earlier work \cite{zhukFDTD}. 

\begin{figure}
\includegraphics[width=1\columnwidth]{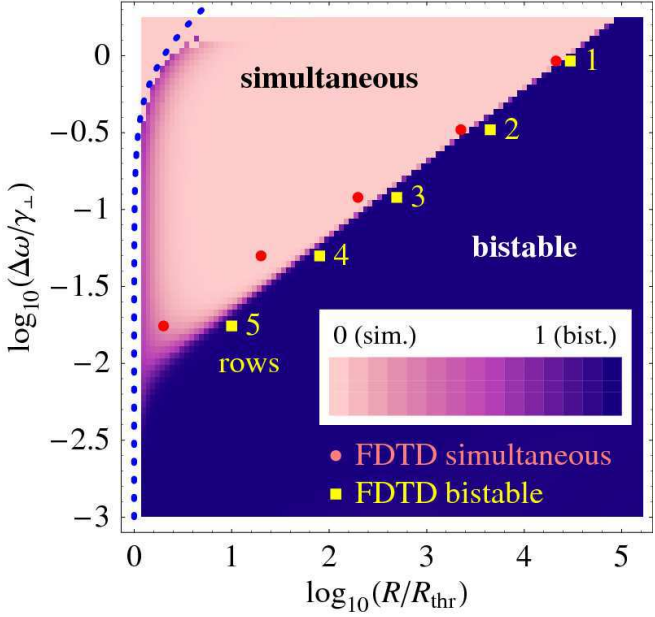}

\caption{(Color online) The dependence of lasing regime on pumping rate~$R$
and intermode frequency separation~$\Delta\omega$ in a class-C laser
model. The parameters are $\kappa_{j}\simeq0.1\gamma_{\perp}$ and
$\gamma_{\parallel}\simeq10^{-4}\gamma_{\perp}$, as used for numerical
simulation in \cite{zhukPRL}. The density plot shows the quantity
$\left|\left|E_{1}(t)\right|-\left|E_{2}(t)\right|\right|/\max\left(\left|E_{1}(t)\right|,\left|E_{2}(t)\right|\right)$
for large $t\gg\gamma_{\perp}^{-1},\gamma_{\parallel}^{-1},\kappa_{j}^{-1}$.
Near-zero (light) values indicate two-mode (simultaneous) lasing while
near-unity (dark) values indicate one-mode (bistable) lasing. The
lasing threshold depending on~$\Delta\omega$ is marked with the
dotted line. Numerical results of the FDTD simulations for coupled-defect
structures (Fig.~\ref{fig:FDTD_structure}) are superimposed over
the density plot. Circles (red) and squares (yellow) show the location
of points where simultaneous and bistable lasing, respectively, was
observed in the mode dynamics during simulations. \label{fig:CLASS_C}}
\end{figure}

More interestingly, an increase of~$R$ can restore bistable lasing
in the cases when simultaneous lasing is observed just above threshold.
Fig.~\ref{fig:GRAPH_C} suggests that there should be no bistability
in the area around $\Delta\omega\simeq\gamma_{\perp}$. The numerical
solution of the class-C equations shows that this is indeed the case
for smaller~$R$. However, if the pumping is increased beyond a certain
critical value~$R_{c}$, a transition from simultaneous to bistable
lasing occurs (Fig.~\ref{fig:CLASS_C}). This effect was reported
earlier \cite{zhukFDTD} with the observation that bistability ensues
when pumping becomes so large that relaxation oscillations disappear.
Our further investigations have revealed that this observation was
rather a coincidence, and $R_{c}$~scales with~$\Delta\omega$ (Fig.~\ref{fig:CLASS_C}),
bifurcating from threshold at approximately the point where $C_{\omega}=1$
according to Eq.~\eqref{eq:C_w}. This falls in line with the result
of the previous section that a stronger pumping is capable of restoring
bistability where it has been deteriorated by adverse effects of insufficient
mode matching.

\begin{figure}
\includegraphics[width=0.75\columnwidth]{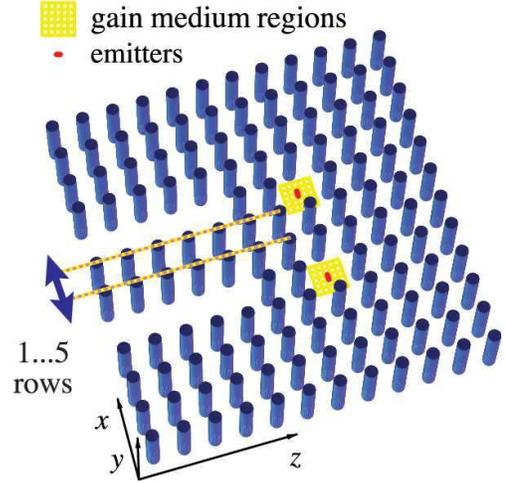}

\caption{(Color online) The family of structures used in numerical FDTD simulations,
based on two coupled defects in a 2D photonic crystal lattice \cite{zhukPRL,zhukFDTD}.
Placing a different number of lattice rows between the defects (1--5),
the intermode frequency separation~$\Delta\omega$ can be changed.
\label{fig:FDTD_structure}}
\end{figure}

Because applicability of the expansion \eqref{eq:decomp_P} and sometimes
even of the SVEA \cite{TureciOE} may become questionable far above
threshold, we have carried out a comparison of Class-B/C results with
direct numerical simulations. As previously described in Ref.~\cite{zhukFDTD},
a space-time FDTD solver was coupled to the four-level laser rate
equations in order to model the response of a laser medium. A 2D photonic
crystal lattice with two coupled defects \cite{zhukPRL} was used
as a model system (Fig.~\ref{fig:FDTD_structure}). Both defects
are filled with four-level gain medium and contain a dipole source
in the centre . By exciting these sources with varying amplitude/phase
relations, the two modes (symmetric and antisymmetric \cite{zhukFDTD})
can be excited in any proportion and thus the initial state of the
resonator can be controlled. By changing the number of lattice rows
between the defects from 1 to 5, one can change~$\Delta\omega$ from~$\sim\gamma_{\perp}$
down to $\sim10^{-2}\gamma_{\perp}$ . The waveguides coupled to the
defects form the primary channel for the radiation to leak out of
the resonator. Care was taken that the mode $Q$-factors remain approximately
the same across the whole family of structures. 

The results of the FDTD simulation runs are superimposed in the phase
diagram in Fig.~\ref{fig:CLASS_C}. For all values of~$\Delta\omega$,
the transition between simultaneous and bistable lasing was found
approximately around~$R_{c}$ as predicted by the analytical theory.
For larger~$\Delta\omega$ the correspondence is better because smaller
$\Delta\omega$~and~$R$ require much longer times to get to the
steady state and there is an increased sensitivity to mode mismatch
(see Fig.~\ref{fig:I12_DOMAIN_R}). Hence it becomes more difficult
to establish the transition point between simultaneous and bistable
lasing with good accuracy. %
\begin{comment}
This is a natural downside to a 2D FDTD scheme applied to microstructured
cavities as the time step has to be much smaller than the slowest
processes in the laser operating slightly above threshold.
\end{comment}
{} 

\begin{figure*}
\includegraphics[width=0.9\textwidth]{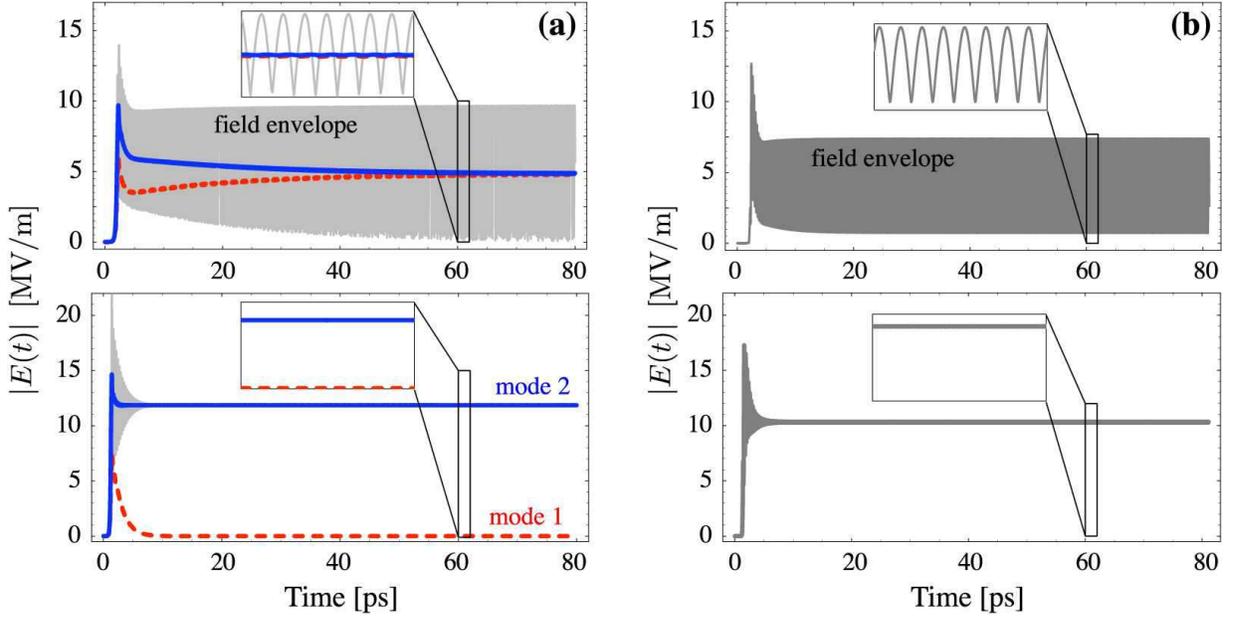}

\caption{(Color online) Comparison between laser field dynamics obtained by
(a) the Class-C equations and (b) direct FDTD numerical simulations
for the 2D PhC structure (see Fig.~\ref{fig:FDTD_structure}) with
defects separated by 2 lattice rows (point \#2 in Fig.~\ref{fig:CLASS_C},
$\log_{10}\left[\Delta\omega/\gamma_{\perp}\right]=-0.481$) for $R<R_{c}$
(\emph{top}) and $R>R_{c}$ (\emph{bottom}). For the Class-C coupled
mode theory results, the red dashed and blue solid lines show the
mode amplitudes $\left|E_{1,2}(t)\right|$, respectively. Initially
both modes are excited and the second mode is given an advantage ($E_{1:2}^{(0)}=2:3$).
The gray line shows the electric field at the mode's maximum $\left|E_{r}(t)\right|=u_{\text{max}}\left|E_{1}(t)\mathrm{e}^{\phi_{+}}+E_{2}(t)\mathrm{e}^{\phi_{-}}\right|$.
For the FDTD results, the field envelope at the center of either defect
$\left|E(\mathbf{r}_{c},t)\right|$ is shown (sampled at local extrema
of light oscillations). The insets show an enlarged portion of the
plots to show the $2\Delta\omega$ intermodal beating whenever both
modes lase at the same time. \label{fig:FDTD_2}}
\end{figure*}

\begin{figure*}
\includegraphics[width=0.9\textwidth]{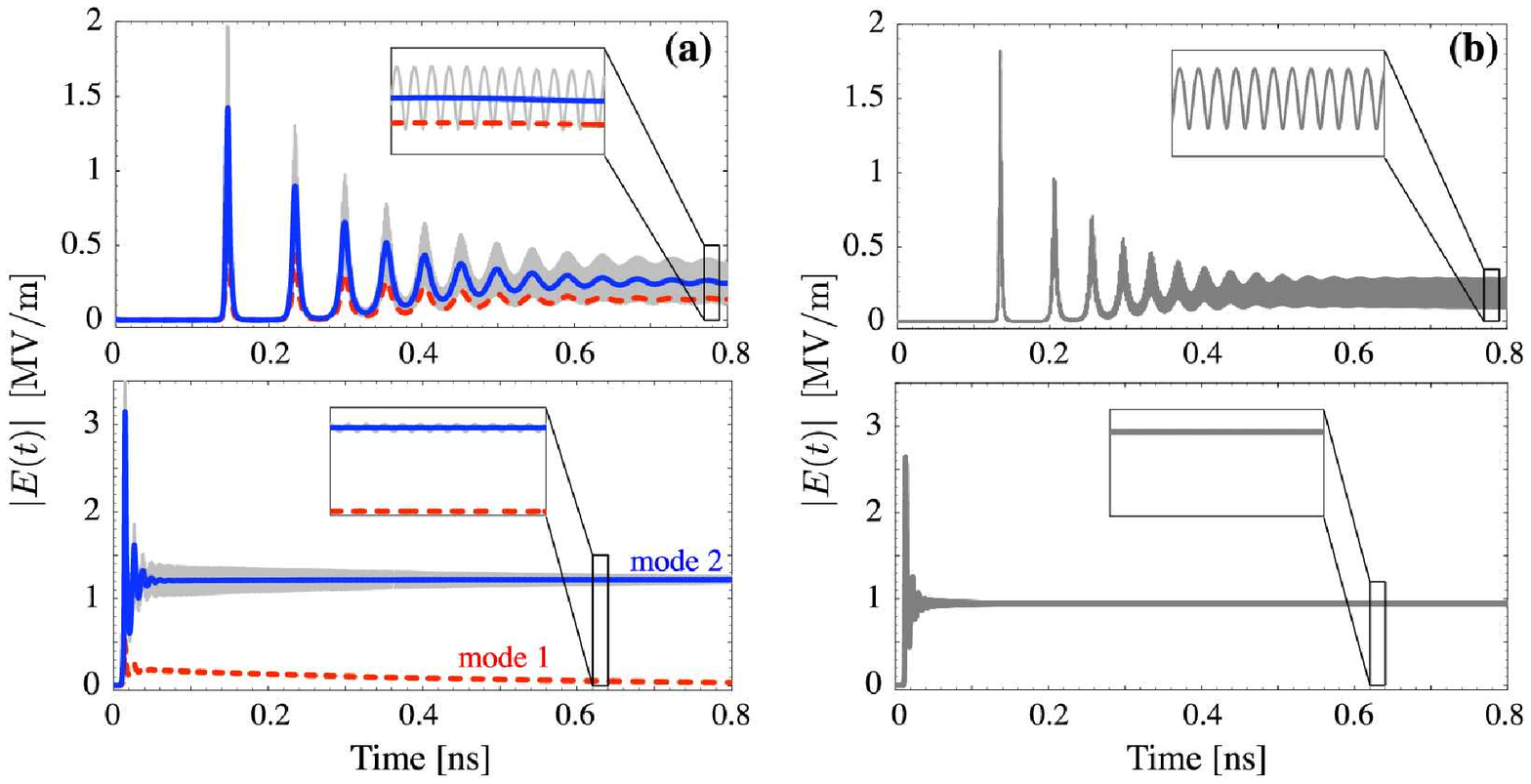}

\caption{(Color online) Same as Fig.~\ref{fig:FDTD_2} but for the 2D PhC
structure with defects separated by 4 rows (point \#4 in Fig.~\ref{fig:CLASS_C},
$\log_{10}\left[\Delta\omega/\gamma_{\perp}\right]=-1.301$). \label{fig:FDTD_4}}
\end{figure*}

In Figs.~\ref{fig:FDTD_2} and~\ref{fig:FDTD_4}, temporal laser
dynamics in numerical simulations and the Class-C model are compared.
We analyze the electric field in the center of either defect~$\mathbf{r}_{c}$.
For FDTD, it is monitored directly by recording the field at the corresponding
point in space $E(\mathbf{r}_{c},t)$. To reduce the excessive amount
of data, we sample the field only at the local maxima, so that an
envelope over the light oscillations is plotted. In the case of the
coupled mode theory, the same quantity is obtained from the mode amplitudes~$E_{1,2}(t)$
using Eq.~\eqref{eq:decomp_E} as $E_{r}(t)=u_{\text{max}}\left(E_{1}(t)\mathrm{e}^{-\mathrm{i}\omega_{1}t}+E_{2}(t)\mathrm{e}^{-\mathrm{i}\omega_{2}t}\right)$
where $u_{\text{max}}=\max\left[u(\mathbf{r})\right]$, assuming the
modes are normalized according to Eq.~\eqref{eq:orthogonal_cavity}.
Taking the absolute value, light oscillations are also neglected,
so the results can be compared to the simulations. 

In all examples of Figs.~\ref{fig:FDTD_2},~\ref{fig:FDTD_4} (which
correspond to the laser operating way above threshold), the field
dynamics shows a good qualitative and quantitative correspondence.
Below~$R_{c}$ where simultaneous two-mode lasing is expected, the
in-cavity field envelope shows the characteristic $2\Delta\omega$
beat oscillations (see the insets in Figs.~\ref{fig:FDTD_2}--\ref{fig:FDTD_4}),
marking the presence of both modes in the laser radiation. Above~$R_{c}$,
the steady-state envelope is flat, indicative of single-mode lasing,
and the beat oscillations are seen to vanish. This corresponds to
quenching of the weaker mode in agreement with theoretical expectations
in the bistable regime. 

Some quantitative discrepancies between the model and simulation results
can be noticed. Some of them (e.g., temporal shifts of the spikes
in Fig.~\ref{fig:FDTD_4}) result from minor deviations in parameters
between the real simulated structure and an idealized two-mode system
considered. These deviations can be compensated for by fine-tuning
the model \cite{zhukFDTD}. Other discrepancies such as the the difference
in the field amplitudes (both at spike maxima and in the steady-state)
can be attributed to gain saturation, which may introduce correction
to the form of the expansion~\eqref{eq:decomp_P} for the gain medum
polarization. %
\begin{comment}
As expected, the deviation between coupled-mode and FDTD amplitudes
is larger for stronger pumping because Eq.~\eqref{eq:decomp_P} gradually
becomes invalid as the pumping rate is increased well above threshold.
\end{comment}
{}This is a limitation inherent in the present coupled-mode model. However,
Eqs.~\eqref{eq:Ec}--\eqref{eq:Wc} and \eqref{eq:Eb}--\eqref{eq:Wb}
are clearly seen to provide a valid description of laser mode dynamics
scenario for relatively strong pumping, unlike the near-threshold
(third-order nonlinearity) theories which are reported to fail badly
in this regime (see \cite{Tureci06}), just like the Class-A equations
\eqref{eq:Ea_final} would. One can overcome this limitation, e.g.,
following the approach in Refs.~\cite{Tureci06,Tureci07} where a
generalization of Eqs.~\eqref{eq:decomp_E}--\eqref{eq:decomp_P}
is introduced. A very good agreement with numerics is reported recently
\cite{TureciOE}. However, only the time-independent (steady-state)
theory is formulated so far. 

The knowledge that stronger pumping can restore a laser into the bistable
regime for higher~$\Delta\omega$ is important in the design of a
laser that can have its wavelength switched by a large value (such
as several tens of nanometers in Refs.~\cite{zhukPRL,zhukPSS}) .
A rigorous explanation of this result is yet to be given. Intuitively,
stronger pumping rates cause shorter lasing onset times compared to
the cavity round-trip time, so the domination of the stronger mode
can occur before the modes have a chance to balance themselves through
the cavity. Indeed, it could be noticed that the transition from simultaneous
to bistable lasing around~$R_{c}$ is accompanied by the disappearance
of $4\Delta\omega$-pulsations in the phase of some dynamical variables.
This suggests that shorter onset due to stronger pumping allows some
of the variables to become phase-locked, which in turn influences
the whole character of the mode interaction (as was seen when the
transition from Eq.~\eqref{eq:Ea_relaxed} to Eq.~\eqref{eq:Ea_final}
was discussed). The detailed investigation of this effect is a subject
for further studies.

\section{Conclusions and outlook\label{sec:SUMMARY}}

In this work, we have addressed the problem of bistability in a microlaser
by systematically formulating the coupled-mode model without prior
assumptions on the mode or cavity geometry, other than the requirement
of the mode orthogonality in the cavity as well as in the gain region
as described by Eqs.~\eqref{eq:orthogonal_cavity} and~\eqref{eq:orthogonal_gain}.
The governing equations have been derived for all laser classes, Eqs.~\eqref{eq:Ec}--\eqref{eq:Wc},
\eqref{eq:Eb}--\eqref{eq:Wb}, and \eqref{eq:Ea_relaxed} for class-C,~B,
and~A, respectively. The issue of classifying the laser dynamics
in the multimode case has been revisited taking into account the intermode
frequency separation~$\Delta\omega$ as a new parameter influencing
the laser dynamics. The model has been derived for the case of two
modes; however, its extension to the case of several modes can be
performed along the same lines. 

The simplest case of the class-A laser equations has been analytically
investigated. It has been shown that coherent mode interaction processes
(population pulsations) can provide an additional mode coupling channel
besides incoherent mode interaction (spatial hole burning). This additional
coupling is what brings the laser into the bistable regime, allowing
the lasing mode to be chosen on demand by the initial condition of
the cavity. This result agrees with the early theoretical predictions
\cite{Siegman,LambLaser}. However, microcavity modes can have a far
better matched intensity distribution inside the gain region, see
Eq.~\eqref{eq:intensities}, compared to bulk-cavity modes, which
are usually heavily out of match unless~$\Delta\omega=0$. As such,
only a moderate amount of pulsation-induced mode coupling is enough
to enter the bistability regime in the case of a microlaser. This
means that microlasers can be bistable in a far greater parameter
range than bulk-cavity lasers, e.g., for much larger~$\Delta\omega$
(Fig.~\ref{fig:GRAPH_C}). We have also shown that a sizable mismatch
in the system parameters that favors one of the modes can destroy
any chance of bistable operation. However, a mismatch with respect
to one parameter can be compensated for by a mismatch with respect
to another (Fig.~\ref{fig:I12_ZERO}). Again, due to better matched
intensity distributions of microcavity modes the bistable regime is
more tolerant to such perturbations (Fig.~\ref{fig:I12_DOMAIN_C}). 

In the more general class-B or class-C laser models, we have shown
numerically that even when $\Delta\omega$~is too large to allow
bistability in the near-threshold class-A case, it can be overcome
by increasing the pumping rate (Fig.~\ref{fig:CLASS_C}). The results
of the theory are confirmed by direct numerical FDTD simulations and
are shown to be qualitatively valid for pumping rates several orders
of magnitudes above the lasing threshold. Further results on bistability
in class-B/C microlasers will be available in a forthcoming publication.

Bistable operation of a multimode microlaser can be useful in many
respects. Since there is no need for an external (and potentially
slow) cavity-tuning process, ultrafast all-optical mode switching
mechanisms can be imagined. Such switching, occurring across $\simeq20$~nm
on a picosecond time scale had indeed been demonstrated numerically
in our earlier work \cite{zhukPRL}. The fast switching between stable
states and the relatively low power of microlasers can be used in
the design of an optical memory (flip-flop) cell. We believe, that
a compact-sized microlaser capable of multiple-wavelength operation
in a wide wavelength range can find numerous applications in integrated
optics and optical communication.

\begin{acknowledgments}
The authors would like to thank C.~Kremers for his assistance and
advice on numerical simulation, as well as A.~V.~Lavrinenko for
stimulating discussions. Financial support from the Deutsche Forschungsgemeinschaft
(DFG FOR 557) is gratefully acknowledged.
\end{acknowledgments}


\begin{thebibliography}{20}
\bibitem{VahalaReview}K.~J.~Vahala, Nature \textbf{424}, 839 (2003).

\bibitem{PBGdefect}O.~Painter, R.~K.~Lee, A.~Scherer, A.~Yariv,
D.~O'Brien, P.~D.~Dapkus, and I.~Kim, Science \textbf{284}, 1819
(1999).

\bibitem{PBGlaser}M.~Imada, A.~Chutinan, S.~Noda, and M.~Mochizuki,
 \emph{}Phys.~Rev.~B \textbf{65}, 195306 (2002).

\bibitem{BabaDisks}S.~Ishii and T.~Baba, \emph{}Appl.~Phys.~Lett.
\textbf{87}, 181102 (2005).

\bibitem{NatRings}M.~Hill, H.~Dorren, T.~de~Vries, X.~Leijtens,
J.~Hendrik den Besten, B.~Smalbrugge, Y.-S.~Oei, H.~Binsma, G.-D.~Khoe,
and M.~Smit, Nature \textbf{432}, 206 (2004).

\bibitem{Siegman}A. Siegman, \emph{Lasers} (University Science Books,
Mill Valley, CA, 1986), Ch.~25.4.

\bibitem{LambLaser}M.~Sargent~III, M.~O. Scully, and W.~E.~Lamb,~Jr.,
\emph{Laser Physics} (Addison-Wesley, Reading, MA, 1974), Ch.~9.

\bibitem{diRing}M.~Sorel, P.~J.~R. Laybourn, A.~Scir\`e, S.~Balle,
G.~Giulliani, R.~Miglierina, and S.~Donati, Opt.~Lett. \textbf{27},
1992 (2002).

\bibitem{mcSatAbsorb}C.~L.~Tang, A.~Schremer, and T.~Fujita,
Appl.\emph{~}Phys.\emph{~}Lett. \textbf{51}, 1392 (1987).

\bibitem{mcSatAbsorb2}C.-F.~Lin and P.-C.~Ku, IEEE J.\emph{~}Quant.\emph{~}Electron.
\textbf{32}, 1377 (1996).

\bibitem{diPBLD}H.~Kawaguchi, IEEE J.~Sel.~Top. Quant.~Elecron.
\textbf{3}, 1254 (1997). 

\bibitem{mcCoupledAgrawal}G.~P.~Agrawal and N.~K.~Dutta, J.~Appl.~Phys.
\textbf{56}, 664 (1984).

\bibitem{mcCoupledOudar}R.~Kuszelewicz and J.~L.~Oudar, IEEE J.~Quant.~Electron.
\textbf{QE-23}, 411 (1987).

\bibitem{mcWieczorekPRA}S.~W.~Wieczorek and W.~W.~Chow, Phys.~Rev.~A
\textbf{69}, 033811 (2004).

\bibitem{mcWieczorekOC}S.~W.~Wieczorek and W.~W.~Chow, Opt.\emph{~}Commun.
\textbf{246}, 471 (2004).

\bibitem{diMMI-BLD}M.~Takenaka, K.~Takeda, Y.~Kanema, Y.~Nakano,
M.~Raburn, and T.~Miyahara, Opt.~Express \textbf{14}, 10785 (2006).

\bibitem{zhukPRL}S. V. Zhukovsky, D. N. Chigrin, A. V. Lavrinenko,
and J. Kroha, Phys.~Rev.~Lett. \textbf{99}, 073902 (2007).

\bibitem{zhukPSS}S. V. Zhukovsky, D. N. Chigrin, A. V. Lavrinenko,
and J. Kroha, Phys. Stat. Sol. (b) \textbf{244}, 1211 (2007).

\bibitem{diRingFlipFlop}S.~Zhang, D.~Lenstra, Y.~Liu, H.~Ju,
Z.~Li, G.~D.~Khoe, and H.~J.~S. Dorren, Opt.~Commun. \textbf{210},
85 (2007).

\bibitem{Tureci06}H.~E.~T\"ureci, A.~Douglas Stone, and B.~Collier,
Phys.~Rev.~A \textbf{74}, 043822 (2006).

\bibitem{Tureci07}H.~E.~T\"ureci, A.~Douglas Stone, and Li~Ge,
Phys.~Rev.~A \textbf{76}, 013813 (2007).

\bibitem{TureciScience}H.~E.~T\"ureci, Li~Ge, S.~Rotter, and
A.~Douglas Stone, Science \textbf{320}, 643 (2008).

\bibitem{mcHodges}S. E. Hodges, M. Munroe, J. Cooper, and M. G. Raymer\emph{,}
J.~Opt. Soc.~Am.~B \textbf{14}, 191 (1997).

\bibitem{mcJohnBusch}L. Florescu, K. Busch, and S. John, J.~Opt.
Soc. Am.~B \textbf{19}, 2215 (2002).

\bibitem{zhukFDTD}S. V. Zhukovsky, D. N. Chigrin, Phys. Stat. Sol.
(b) \textbf{244}, 3515 (2007).

\bibitem{mcZehnle}V.~Zehnl\'e, Phys.~Rev.~A \textbf{57}, 629
(1998).

\bibitem{Haken} H. Haken and H. Sauermann, Z. Phys. \textbf{173},
261 (1963); 

\bibitem{HakenFu}H. Fu and H. Haken, Phys. Rev. A \textbf{43}, 2446
(1991),

\bibitem{Peterman}A.~E.~Siegman, Phys.~Rev.~A \textbf{39}, 1253
(1989). 

\bibitem{Svelto}O.~Svelto, \emph{Principles of lasers} (Plenum Press,
New York, 1989), Ch.~6.

\bibitem{TureciOE}Li Ge, R.~J.~Tandy, A.~Douglas Stone, and H.~E.~T\"ureci,
Opt.~Express \textbf{16}, 16895 (2008).
\end{thebibliography}
\end{document}